\documentclass[5p,times]{elsarticle}
\usepackage{amsmath}
\usepackage[colorlinks]{hyperref}
\usepackage{multirow}
\usepackage[numbers]{natbib}
\usepackage{caption}
\usepackage{tikz} \usetikzlibrary{arrows.meta,patterns}

\newcommand{\vect}[1]{\mathbf#1}
\newcommand{\nab}{\vect{\nabla}}

\newcommand{\toprule}{\noalign{\hrule height 2\arrayrulewidth}}
\newcommand{\bottomrule}{\toprule}
\newcommand{\midrule}{\hline}

\begin{document}
\sloppy
\let\WriteBookmarks\relax
\def\floatpagepagefraction{1}
\def\textpagefraction{.001}

\title{Sphere-based test cases for open boundary simulations in\\electrostatics and magnetostatics}                      

\author[1]{Marc Boulé}
\ead{marc.boule@etsmtl.ca}
\affiliation[1]{organization={École de technologie supérieure},
                addressline={1100 Notre-Dame Ouest}, 
                city={Montréal},
                postcode={H3C 1K3}, 
                state={Québec},
                country={Canada}}

\begin{abstract}
This paper presents a library of sphere-based test cases for verifying the correctness of open boundary simulations in electrostatics and magnetostatics for computational physics and engineering applications. While the energy in the fields of a single sphere is computed for various source configurations, the energy, force and torque (when applicable) are also computed for a pair of spheres. An extension to electromagnetic momentum is also given for the single sphere, demonstrating its use as a benchmarking metric. Analytical expressions are provided for each test case, from which the validity of a simulation environment can be assessed. The finite element method is used as the example context, and four different open boundary methods are evaluated. Once verified, the simulation is used to visualize the Poynting vector field associated with an electric charge placed near a permanent magnet, a configuration for which no analytical expression is available. Simulation files are made available and all numerical experiments are reproducible with the ONELAB open source software.
\\ This article appeared in \textit{Computer Physics Communications}, vol. 331, article 110500, Feb. 2027, and may be downloaded for personal use only. 
{\small\url{https://doi.org/10.1016/j.cpc.2026.110500}}
\end{abstract}


\maketitle

\section{Introduction}
Time-invariant electromagnetic fields involve three main regimes~\cite{PIRIOU2024}: electrostatics, magnetostatics and electrokinetics (also called current flow). In the first two regimes, it is often convenient to place the outer boundary at infinity in order to simplify the problem. This can also occur in many other areas such as fluid flow~\cite{BAGANIS2014}, linear elasticity~\cite{KUMAR2000}, acoustics~\cite{NABIZADEH2021}, seismic loading~\cite{ASHEGHABADI2020}, and countless electromagnetic examples including the study of radiating fields~\cite{SUGAHARA2015}, microstrip lines and bus bars~\cite{FREEMAN1988,MEEKER2013}. This paper focuses on electrostatics and magnetostatics and presents a set of sphere-based test cases for verifying the correctness of open boundary simulations in computational physics and engineering. Configurations with exact analytical expressions are sought, in contrast to approximate analytical expressions that have simplifying assumptions. Although the finite element method (FEM) is used in the numerical experiments, results apply to other methods as well, such as the finite difference method (FDM) and the method of moments (MoM)~\cite{JIN2015}.  

Ensuring that a simulation behaves as expected involves verifying multiple aspects: differential equations in the model, boundary conditions, physical parameters, meshing parameters, software settings, etc. In many circumstances, it can be beneficial to run prior simulations involving test cases for which the expected results are available. If exact expressions are known for the fields, pointwise checks and trace evaluations can be performed. When more advanced post-processing integrals can be performed, $L_2$ error norms~\cite{FISH2007} can also be used to evaluate the difference between expected and simulated fields. 

Another popular method is to use the energy in the electric or magnetic fields. 
Energy is also one of the methods used to calculate capacitance, inductance, force and torque, and can be easily evaluated in practically any simulation tool. 
A valid total energy value, computed from the integral of the norm of a vector field, does not by itself guarantee that the simulation is working properly since the field may still have an incorrect direction throughout the domain. Nevertheless, total energy remains a useful and practical global indicator. When electrostatic and magnetostatic fields coexist in the same simulation, electromagnetic momentum is a global vector quantity that does reflect field direction, taking the form of linear or angular momentum, depending on the situation.

The sphere remains relatively underutilized in the open boundary literature, and this work aims to provide a more comprehensive analysis. Particular attention is given to expressions for energy, force, torque and momentum as benchmarking metrics. To the best of the author's knowledge, the momentum results for the sphere (excluding those involving current sources), along with the fields of the permanently polarized or magnetized sphere, comprise generalizations for arbitrary permittivity and/or permeability not previously reported. The majority of the force and torque results are built upon known field or dipole expressions, as are a few of the energy results, while others appear directly in prior work and are included for completeness. The library of test cases presented in Sections~\ref{sec_ele} to \ref{sec_em} serves several purposes:
\begin{enumerate}
\item to evaluate the performance of different boundary methods, including the effect of domain truncation when a true open boundary method cannot be applied;
\item to help verify the correct implementation of simulation codes or the proper configuration of commercial software;
\item to confirm that simulation results are consistent and that errors remain within reasonable bounds before developing more complex models;
\item to assess how different applications report energy and coenergy in post-processing, given the different ways in which these terms are defined in the literature.
\end{enumerate}

Section~\ref{sec_sims} presents FEM experiments using the open source ONELAB software and concludes with a visualization of the Poynting vector field associated with a charge placed near a bar magnet. All simulation files used to generate the reported results are made available to facilitate their reproduction and further exploration~\cite{SIMFILES}.

\section{Background}
\label{secbg}

This section presents a quick overview of open boundary methods and related test cases appearing throughout the literature, and concludes by summarizing the test cases that follow in Sections~\ref{sec_ele} to~\ref{sec_em}.

\subsection{Open boundary methods}
\label{ssecobm}

When modeling problems with infinite domains, many computational methods such as  FEM and FDM require a discretized finite region over which the calculations are performed. When an infinite space must be represented by a finite number of elements or a finite grid, different methods can be used. The simplest method is called truncation, where the outer boundary is placed at a finite distance. This can be implemented in any application; however, it usually results in reduced accuracy.

\begin{figure}
\centerline{

\begin{tikzpicture}
\def\scl{1.0}
\def\rb{2}
\def\rc{0.4}
\fill[fill=gray!20!white] (-4,0) rectangle (-1,2.75);
\fill[fill=gray!20!white] (-4,2.75) .. controls (-2,3.2) and (-1.2,3.1) .. (-1,2.75);
\fill[fill=gray!20!white] (-1,2.75) .. controls (-0.6,2.05) and (-0.8,0.5) .. (-1,0);
\draw[fill=white      ] (-4,0) -- (-2,0) arc[start angle=0, end angle=90, radius=\rb] --cycle;
\draw[-{Stealth[scale=1.4]}] (-4,0) -- (-0.5,0) node[anchor=south] {\scalebox{\scl}{$y$}};
\draw[-{Stealth[scale=1.4]}] (-4.0,0) -- (-4,3.25) node[anchor=west] {\scalebox{\scl}{$z$}};
\draw[densely dashed] (-4,0.5) arc[start angle=-90, end angle=90, radius=\rc];
\coordinate (P) at (-2.5,2.25);
\fill[] (P) circle (0.06) node[anchor=west] {\scalebox{\scl}{\hspace{4pt}$p=(r,\theta,\phi)$}};
\draw[] (-3.25,1.45) node[] {\scalebox{\scl}{$r$}};
\draw[densely dotted] (-4,0) -- (P);
\draw[] (-2.7,0.95) node[] {\scalebox{\scl}{$r_b$}};
\draw[densely dotted] (-4,0) -- ++(28:2);

\draw[fill=gray!20!white] (0.5,0) -- (3.3,0) arc[start angle=0, end angle=90, radius=2.8] --cycle;
\draw[fill=white] (0.5,0) -- (2.5,0) arc[start angle=0, end angle=90, radius=\rb] --cycle;
\draw[-{Stealth[scale=1.4]}] (0.5,0) -- (4.0,0) node[anchor=south] {\scalebox{\scl}{${y}$}};
\draw[-{Stealth[scale=1.4]}] (0.5,0) -- (0.5,3.25) node[anchor=west] {\scalebox{\scl}{${z}$}};
\draw[densely dashed] (0.5,0.5) arc[start angle=-90, end angle=90, radius=\rc];
\coordinate (PT) at (1.725,1.837);
\fill[] (PT) circle (0.06) node[anchor=west] {\scalebox{\scl}{\hspace{5pt}$\tilde{p}=(\tilde{r},\theta,\phi)$}};
\draw[densely dotted] (0.5,0) -- (PT);
\draw[] (1.2,1.4) node[] {\scalebox{\scl}{$\tilde{r}$}};
\draw[densely dotted] (0.5,0) -- (PT);
\draw[] (1.8,0.95) node[] {\scalebox{\scl}{$r_b$}};
\draw[densely dotted] (0.5,0) -- ++(28:2);
\draw[] (2.7,0.60) node[] {\scalebox{\scl}{$r_e$}};
\draw[densely dotted] (0.5,0) -- ++(10:2.8);

\draw [->] (-2.4,2.4) to [bend left=40] (1.65, 1.95);
\end{tikzpicture}

}
\caption{Transformation of a point outside of the inner boundary (left) into the shell region (right). An uncentered sphere (dashed) is the object of simulation. \label{figShell}}
\end{figure}
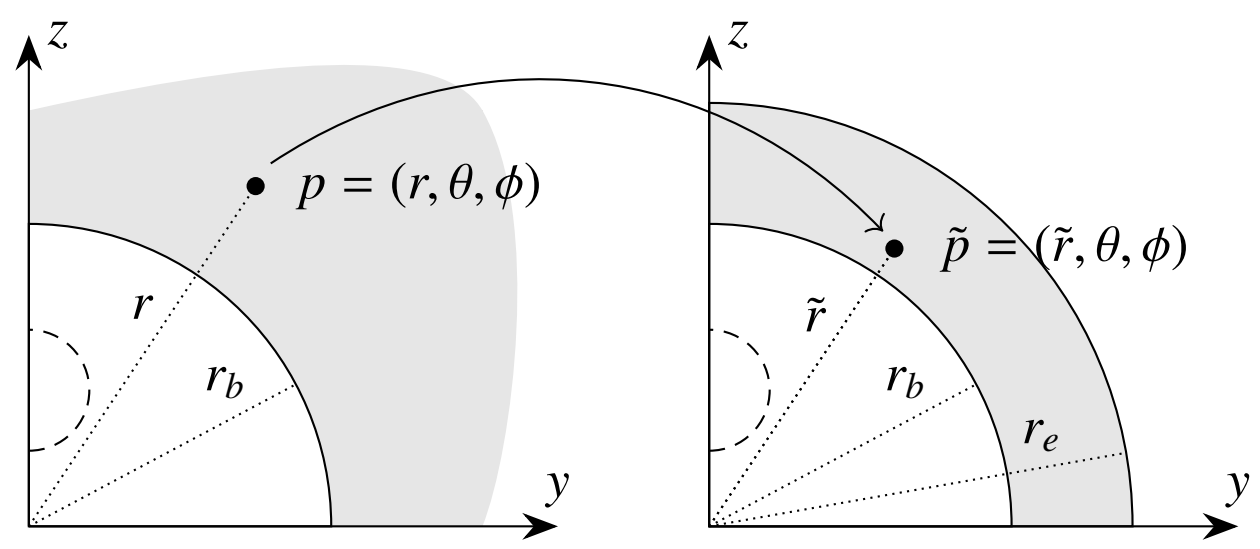

The shell method~\cite{HENROTTE1999} is an exact approach based on a spatial transformation. It can be used to map the infinite space outside a spherical boundary into a spherical shell. The radius of the untransformed inner region is $r_b$ and the outer radius of the shell is $r_e$. The shell method can also be applied to cylindrical domains~\cite{BOULE2026}. 
In spherical coordinates (see Fig.~\ref{figShell}), the angles $\theta$ and $\phi$ are unaffected, while $r$ is transformed according to
\begin{equation}
\label{eqshell}
\tilde{r}= 
  \begin{cases}
    r & \textrm{if}~ r\leq r_b\\
    r_e-{r_b(r_e- r_b)}/{r} & \textrm{if}~ r>r_b
  \end{cases}
\end{equation}

In the ABC method (asymptotic boundary condition), the outer region is entirely removed and an approximate matching is created on the boundary~\cite{CHEN1994,GRATKOWSKI2000,GRATKOWSKI2002}. First order ABCs requires the use of a non-homogeneous Neumann boundary condition, which translates to a potential-dependent surface charge or current density on this outer boundary. The following terms are used in the weak formulations~\cite{PIRIOU2024} to implement the first order ABC:
\begin{equation}
\label{eqabc}
\vect{n}\cdot \vect{D} = \frac{m\varepsilon_0V}{r_e},\qquad \vect{n}\times \vect{H} =  \frac{\vect{A}_t}{\mu_0 r_e},
\end{equation}
where~$\vect{n}$ is the unit vector normal to the boundary (pointing outward), $r_e$ is the radius of this outer boundary (coinciding with the outer radius of the shell method here), $V$ is the electric scalar potential, and $\vect{A}_t$ is the tangential component of the magnetic vector potential. These potentials are part of the unknowns in the simulation. For the electric potential, the integer $m$ is set to~1 when the far field is monopolar, and to~2 when it is dipolar. The first equality above can also be used in a magnetic scalar potential formulation with the substitutions $\vect{D}\rightarrow\vect{H}$, $\varepsilon_0\rightarrow\mu_0$ and $m\rightarrow2$ (no magnetic monopoles). The shell and ABC methods may not be applicable in all software since they usually require more specialized features (the use of Jacobian matrices and boundary terms in FEM formulations, for example).

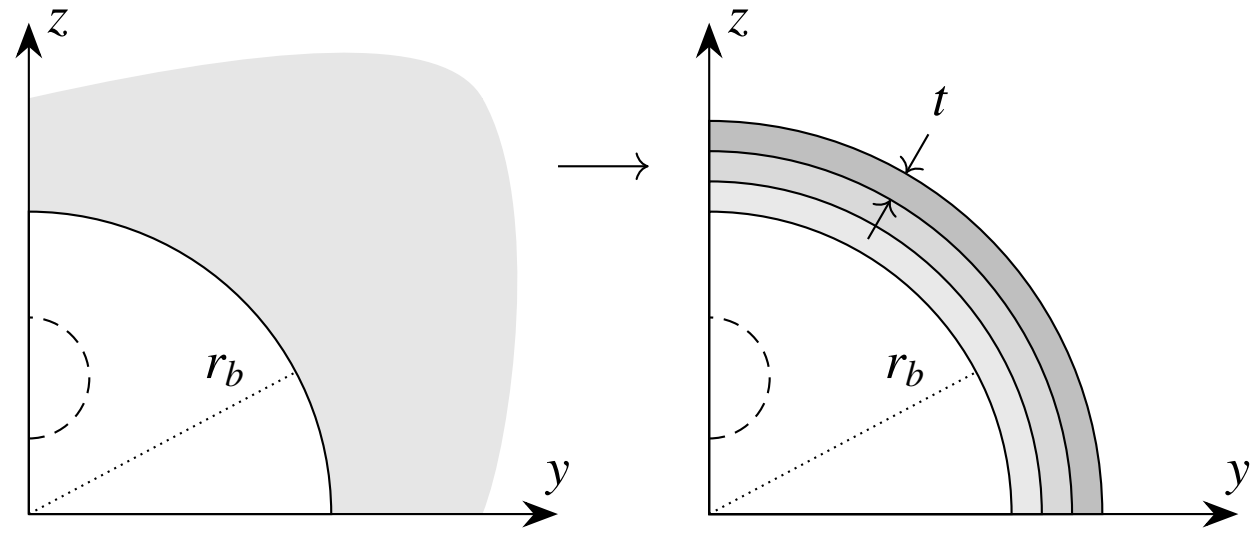
\begin{figure}
\centerline{

\begin{tikzpicture}
\def\scl{1.0}
\def\rb{2}
\def\rc{0.4}
\fill[fill=gray!20!white] (-4,0) rectangle (-1,2.75);
\fill[fill=gray!20!white] (-4,2.75) .. controls (-2,3.2) and (-1.2,3.1) .. (-1,2.75);
\fill[fill=gray!20!white] (-1,2.75) .. controls (-0.6,2.05) and (-0.8,0.5) .. (-1,0);
\draw[fill=white      ] (-4,0) -- (-2,0) arc[start angle=0, end angle=90, radius=\rb] --cycle;
\draw[-{Stealth[scale=1.4]}] (-4,0) -- (-0.5,0) node[anchor=south] {\scalebox{\scl}{$y$}};
\draw[-{Stealth[scale=1.4]}] (-4.0,0) -- (-4,3.25) node[anchor=west] {\scalebox{\scl}{$z$}};
\draw[densely dashed] (-4,0.5) arc[start angle=-90, end angle=90, radius=\rc];
\draw[] (-2.7,0.95) node[] {\scalebox{\scl}{$r_b$}};
\draw[densely dotted] (-4,0) -- ++(28:2);

\draw[fill=gray!50!white] (0.5,0) -- (3.1,0) arc[start angle=0, end angle=90, radius=2.6] --cycle;
\draw[fill=gray!30!white] (0.5,0) -- (2.9,0) arc[start angle=0, end angle=90, radius=2.4] --cycle;
\draw[fill=gray!17!white] (0.5,0) -- (2.7,0) arc[start angle=0, end angle=90, radius=2.2] --cycle;
\draw[fill=white] (0.5,0) -- (2.5,0) arc[start angle=0, end angle=90, radius=\rb] --cycle;
\draw[-{Stealth[scale=1.4]}] (0.5,0) -- (4.0,0) node[anchor=south] {\scalebox{\scl}{${y}$}};
\draw[-{Stealth[scale=1.4]}] (0.5,0) -- (0.5,3.25) node[anchor=west] {\scalebox{\scl}{${z}$}};
\draw[densely dashed] (0.5,0.5) arc[start angle=-90, end angle=90, radius=\rc];
\draw[] (1.8,0.95) node[] {\scalebox{\scl}{$r_b$}};
\draw[densely dotted] (0.5,0) -- ++(28:2);
\draw [->] (1.55,1.8187) to [bend left=0] (1.7,2.0785);
\draw [->] (1.95,2.5115) to [bend left=0] (1.8,2.2517);
\draw[] (1.85,2.5115) node[anchor=south west] {\scalebox{\scl}{$t$}};

\draw [->] (-0.5,2.3) to [bend left=0] (0.1, 2.3);
\end{tikzpicture}

}
\caption{Concentric shells (layers) of a third-order IABC, each of thickness $t$.\label{figIABC}}
\end{figure}

An ad-hoc realization of the ABC (called improvised-ABC, or IABC) was also developed~\cite{MEEKER2013,MEEKER2014}, whereby the approximate matching is accomplished by a series of concentric shells, each having different material properties (relative permittivity or permeability). To avoid confusion with the shell method, these shells are instead referred to as layers (see Fig.~\ref{figIABC}). Each layer has a thickness $t$, and the relative permittivities or permeabilities of the layers depend on the ratio $\delta=t/r_b$. 
The material properties of the layers also depend on the type of boundary condition at infinity (Dirichlet or Neumann) and on the symmetry of the simulation (planar/cylindrical versus axisymmetric/spherical). For example, in a spherical electrostatic potential simulation with a Dirichlet outer boundary ($V=0$), the relative permittivities of the successive layers in a 3rd order IABC with $\delta=0.1$ are, from inner to outer: 0.429378, 10.3276, 0.0820837~\cite{MEEKER2014}. 

Although there is no apparent correspondence for mapping a point in the original exterior region to a point in the IABC layers, a volume integral that includes the layers can be used to compute a proper field energy (electric or magnetic). This property does not apply to electromagnetic momentum, where the pointwise correspondence between electric and magnetic fields does not generally hold in the layers. 
The IABC method can be used in virtually any FEM simulation environment. Although approximate, this method can achieve high accuracy, as demonstrated by the numerical experiments in Section~\ref{sec_sims}. Other methods such as ballooning, infinite elements and the Kelvin transformation (a shell transformation with $r_e=0$) also exist~\cite{CHEN1997}, but are not considered here.

\subsection{Review of prior test cases}

Different test cases have been used in the study of problems involving open boundaries. 
A single conducting cylinder (wire) is of limited interest since its potential difference with respect to infinity diverges. For a pair of infinitely long parallel cylinders with opposite charges and currents (called wirepair), the physical relevance is much greater since the potential decays more rapidly. Yet, it can be shown that when resistance is taken into account, the total energy in the electric field outside of the conductors diverges, even for a segment of finite length along the wires. A volume integral of the electric field in Eq.~(50) in~\cite{BOULE2024} reveals this. 

For a simulation with a cylindrical domain, the $LC$ parameters of the wirepair can nonetheless be used for benchmarking~\cite{BOULE2026}. Comparisons to exact solutions can be performed for cylindrical conductors~\cite{CHEN1994,MEEKER2014,LOWTHER1989,GOLD2020}, while other forms of transmission lines have approximate solutions~\cite{FREEMAN1988,GARCIA2024}. The forces in a three-wire configuration~\cite{SPASOV2022} and the magnetic field of a four-wire configuration~\cite{REMACLE1994} were compared to exact solutions, although no analytical expressions were provided for the magnetic energy.

One of the simplest benchmarking problems is the single conducting sphere in an unbounded vacuum. An electrostatic simulation with a gradual displacement of the sphere's center position was used to evaluate the effectiveness of the IABC method~\cite{MEEKER2014}. 
Another conceptually simple object is the point charge, sometimes modeled with the delta function, but is difficult to implement in FEM/FDM given the diverging fields and potentials at its location.

Exact expressions for the capacitance~\cite{LEKNER2011,BANERJEE2021} and potential~\cite{MORSE1953} of a pair of conducting spheres are known, which is equivalent to a single sphere near a ground plane; however, these involve infinite sums. 
%
Other configurations with either approximate or non-closed forms include coils with circular~\cite{BRAURER1982} or rectangular~\cite{MEEKER2013,FREEMAN1989} cross sections, a thin conducting sheet~\cite{GRATKOWSKI1992}, a square plate capacitor~\cite{THATCHER1978}, a torus~\cite{WONG1985} and two rectangular bus bars~\cite{NOGUEIRA2010}. In some open boundary problems, comparisons to measured values were even performed~\cite{HENROTTE1999,JAISWAL2006}.

It is also possible to devise test cases in which objects are placed in uniform external fields. Two examples are the torque on a rectangular bar magnet~\cite{MEEKER2014} and the fields of a permeable sphere~\cite{BRUNOTTE1992}, both of which have analytical solutions. Such cases are not considered here for consistency with the notion of unbounded space, where a uniform field would imply an infinite energy. A conducting sphere in an infinitely extended homogeneous dielectric~\cite{CIRIC1986,VENKATESHA2000} is also not treated.

A detailed analysis of benchmarking in electrostatics and magnetostatics is presented by Uzunbajakau~\cite{UZUN2024,CEMLOGBOOK}. Although many examples are based on cylindrical geometries and do not require open boundaries, several cases involving spherical shells are presented. The emphasis in that work is placed on the $L_2$ error norm, whereas other quantities such as energy and force are not evaluated.

\subsection{Overview of proposed test cases}

The proposed test cases are based on either the single sphere or a pair of non-intersecting spheres. 
A solid sphere has a relative permittivity $\varepsilon_r$ and a relative permeability $\mu_r$, both of which are uniform and scalar in nature. Vacuum ($\varepsilon_0$, $\mu_0$) surrounds the sphere(s). 
The radius of a sphere is~$R$, its volume region is denoted $\Omega_s$ and its surface is $\Gamma_s$. When two spheres are present, their distance from center-to-center is~$d$. The electric and magnetic sources used in the proposed test cases are:

\begin{enumerate}
\item ${P}_{\!p}$: permanently polarized non-conducting sphere, with given uniform electric dipole moment per unit volume;

\item $\rho_{\!f}$: non-conducting uniformly charged sphere, with given volume free charge density;

\item $V_0$: conducting sphere, with given electric potential; 

\item $\sigma_{\!f}$: sphere (conducting or not), with given uniform surface free charge density; 

\item ${M}_{\!p}$: permanently magnetized sphere (conducting or not), with given uniform magnetic dipole moment per unit volume; 

\item ${J}_{\!f}$: non-conducting sphere of volume charge density $\rho_{\!f}$, spinning with angular velocity $\boldsymbol{\omega}$, with $\varepsilon_r=\mu_r=1$;

\item ${K}_{\!f}$: non-conducting sphere of surface charge density $\sigma_{\!f}$, spinning with angular velocity $\boldsymbol{\omega}$, with $\varepsilon_r=\mu_r=1$. 
\end{enumerate}

The first four cases concern electrostatics, while cases~5 to~7 relate to magnetostatics. Cases~6 and~7 also involve electromagnetic momentum, as does the combination of permanent magnetization with any of the electrostatic sources. The free volume and surface current densities of a rotating sphere are
\begin{equation}
\label{eqJK}
\vect{J}_{\!f}=\rho_{\!f} \boldsymbol{\omega}\times \vect{r},\qquad\vect{K}_{\!f}=\sigma_{\!f} \boldsymbol{\omega}\times \vect{r},
\end{equation}
with $\vect{r}$ in $\Omega_s$ or $\Gamma_s$ respectively. When the sphere is not centered, the substitution $\vect{r}\rightarrow \vect{r}-\vect{r}_c$ can be applied to~(\ref{eqJK}), where $\vect{r}_c$ is the position of the sphere's center.

Two types of integrals can be used to calculate energy and momentum: open space field integrals are general and require integration over the entire domain $\Omega$ (which includes the sphere(s)), while finite space integrals are more specific depending on the chosen source configuration, and only require integration over a small volume or surface (related to the sphere(s)). Finite space integrals are required when integration in the exterior region is not possible. Examples of this are when the ABC method is used, since the exterior region is not present, or when the IABC method is used in a momentum simulation.

In the next four sections, the rightmost column of each table presents analytical expressions for energy, force, torque and momentum for the sphere-based benchmarks. These expressions constitute the central benchmark cases in this work. General field integrals are also given in the tables, and can be used to calculate these quantities for any shape of object. 
With the exception of one test case, electric and magnetic field expressions are provided so that error norms can be computed. Although the analytical calculations for the single sphere are undertaken with its center at the origin, and with polarization, magnetization and angular velocity directed in the $z$-axis, they are subsequently made coordinate independent in their final representations.

\section{Electrostatic energy test cases}
\label{sec_ele}

For a linearly polarizable dielectric with a permanent polarization $\vect{P}_{\!p}$ and a permittivity $\varepsilon=\varepsilon_0\varepsilon_r$, the polarization is~\cite{SESSLER1987}
\begin{equation}
\label{eqPtot}
\vect{P}=\varepsilon_0(\varepsilon_r-1)\vect{E}+\vect{P}_{\!p}
\end{equation}
and the constitutive relation is~\cite{BERE2012}
\begin{equation}
\label{eqCRes}
\vect{D}=\varepsilon_0\vect{E}+\vect{P}=\varepsilon\vect{E}+\vect{P}_{\!p}
\end{equation}
where $\vect{D}$ is the electric displacement and $\vect{E}$ is the electric field. 
In this context, two different energy calculations are possible. To remain consistent with the terminology adopted later in Section~\ref{sec_mag} (magnetostatic energy), the $W_E$ formulas in Table~\ref{tab_es} calculate the energy, while the $W_D$ formulas calculate the coenergy. In a linear dielectric without permanent polarization, energy and coenergy are equal.

\begin{table*}
\centering
\caption{Electrostatic energy test cases for a single sphere (first four sources) and two spheres (last two sources).\label{tab_es}}
\renewcommand{\arraystretch}{1.5}
\begin{tabular*}{0.88\linewidth}{@{\extracolsep{\fill}} llllll@{}}

\toprule
Source & Material & Energy & Open Space & Finite Space & Result for Sphere(s) \\ \midrule

\multirow{2}{*}{${P}_{\!p}$} &
\multirow{2}{*}{Nonconduct. (NC)} & 
$W_E$ & 
$\frac{1}{2}\int_\Omega\varepsilon E^2\,d\Omega$ &   
$-\frac{1}{2}\int_{\Omega_s} \vect{P}_{\!p}\cdot\vect{E}\,d\Omega_s$ & 
$\frac{2\pi}{3\varepsilon_0(\varepsilon_r+2)} R^3 P_{\!p}^2$  \\

&  & 
$W_D$ &
$\frac{1}{2}\int_\Omega\frac{1}{\varepsilon} D^2\,d\Omega$ &
$\frac{1}{2}\int_{\Omega_s}\frac{1}{\varepsilon} \vect{P}_{\!p}\cdot\vect{D}\,d\Omega_s$ & 
$\frac{4\pi}{3\varepsilon_0\varepsilon_r(\varepsilon_r+2)} R^3 P_{\!p}^2$\\[2pt]
\hline 

$\rho_{\!f}$ &
NC & 
& 
&
$\frac{1}{2}\int_{\Omega_s} \rho_{\!f} V\,d\Omega_s$ & 
$\frac{2\pi}{9\varepsilon_0}\left(\frac{1}{5\varepsilon_r}+1\right) R^5\rho_{\!f}^2 $ \\[1pt]
\cline{1-2}\cline{5-6}

$V_0$ &
Conducting (C) & 
$W_E=W_D$ &
$ \frac{1}{2}\int_\Omega\vect{E}\cdot\vect{D}\,d\Omega$ &
\multirow{2}{*}{$\frac{1}{2}\int_{\Gamma_s} \sigma_{\!f} V\,d\Gamma_s$} &
$2\pi\varepsilon_0 R\,V_0^2$\\[1pt]
\cline{1-2}\cline{6-6}

$\sigma_{\!f}$ &
NC ($\varepsilon_r$-ind.) or C & 
 & 
 &
 & 
$\frac{2\pi}{\varepsilon_0} R^3\sigma_{\!f}^2$\\[1pt]
\hline

$\rho_{\!f}\,;\,-\rho_{\!f}$ &
NC; $\varepsilon_r=1$ & 
\multirow{4}{*}{$W_E=W_D$} & 
\multirow{4}{*}{$ \frac{1}{2}\int_\Omega\vect{E}\cdot\vect{D}\,d\Omega$} &
$\frac{1}{2}\int_{\Omega_{s1}\cup\Omega_{s2}} \rho_{\!f} V\,d\Omega_s$ & 
$\frac{4\pi}{9\varepsilon_0}\left(\frac{6}{5}-\frac{R}{d}\right) R^5\rho_{\!f}^2 $ \\[1pt]
\cline{1-2}\cline{5-6}

&
& 
&
&
&
$8\pi\varepsilon_0 R\,V_0^2 \beta\sum_{i=0}^\infty w(i)\,$,~with\\

$V_0\,;\,-V_0$ &
Conducting & 
&
&
$\frac{1}{2}\int_{\Gamma_{s1}\cup\Gamma_{s2}} \sigma_{\!f} V\,d\Gamma_s$ &
$w(i)=((k+\beta)^{2i+1}-1)^{-1},$\\

&
& 
&
&
&
$k=\frac{d}{2R},~\beta=\sqrt{k^2-1}$\\

\bottomrule

\end{tabular*}
\end{table*}

\subsection{Permanent polarization $P_{\!p}$}
\label{ssec_pp}

The electric scalar potentials inside and outside of the permanently polarized sphere are calculated by solving the Laplace equation by separation of variables, making use of Legendre polynomials. The solution can be derived similarly to the spherical linear dielectric placed in a uniform external field~\cite{GRIFFITHS2024}, 
by using the following boundary conditions instead:
\begin{align}
V_\textrm{in} =V_\textrm{out} &\qquad \textrm{at~}r=R\label{eqBCes1}\\
-\varepsilon_0\varepsilon_r\frac{\partial V_\textrm{in}}{\partial r} +P_{\!p}\cos(\theta) = -\varepsilon_0\frac{\partial V_\textrm{out}}{\partial r} &\qquad \textrm{at~}r=R\label{eqBCes2}\\
V_\textrm{out}\rightarrow 0 &\qquad \textrm{for~}r\gg R\label{eqBCes3}
\end{align}
The polarization is in the positive $z$-axis direction, corresponding to $\theta=0$ in spherical coordinates. The second equation above is the boundary condition $D_{\perp\textrm{in}}=D_{\perp\textrm{out}}$ at the surface of the sphere, along with the use of~(\ref{eqCRes}).  Equation~(\ref{eqCRes}) reduces to $\vect{D}=\varepsilon_0\vect{E}$ outside the sphere.

Once solved, the potentials can be used to calculate the $\vect{E}$ fields in both regions with $\vect{E}=-\nab V$ (using spherical coordinates). When generalized for an arbitrary polarization direction and expressed in a coordinate-independent form, the potentials are
\begin{equation}
\label{eqVP}
V_\textrm{in}(\vect{r}) = \frac{\vect{P}_{\!p}\!\cdot\vect{r}}{\varepsilon_0(\varepsilon_r+2)},\qquad\
V_\textrm{out}(\vect{r}) = \frac{R^3}{\varepsilon_0(\varepsilon_r+2)r^3}\,\vect{P}_{\!p}\!\cdot\vect{r},
\end{equation}
and the electric fields are
\begin{equation}
\label{eqEP}
\begin{aligned}
&\vect{E}_\textrm{in}(\vect{r}) =\frac{-\vect{P}_{\!p}}{\varepsilon_0(\varepsilon_r+2)},\\
&\vect{E}_\textrm{out}(\vect{r}) = \frac{R^3}{\varepsilon_0(\varepsilon_r+2)r^3}\left(\frac{3}{r^2}\left(\vect{P}_{\!p}\!\cdot\vect{r}\right)\vect{r}-\vect{P}_{\!p}\right),
\end{aligned}
\end{equation}
with the shorthand $r=\| \vect{r} \|$ assumed throughout the paper. 
The $\vect{D}$ fields then follow naturally from~(\ref{eqCRes}). 
If the sphere is not centered at the origin, the potentials and fields can be evaluated using $V(\vect{r}-\vect{r}_c)$ and $\vect{E}(\vect{r}-\vect{r}_c)$. The electric field is uniform inside the sphere and dipolar outside of the sphere. 

The energy calculations are performed using either the open space or finite space field integrals shown in row ${P}_{\!p}$ of Table~\ref{tab_es}. These volume integrals are most easily evaluated in spherical coordinates (with the sphere centered at the origin). 
The results for the sphere are also shown in the same row of this table.

\subsection{Volume charge density $\rho_{\!f}$}

For a non-conducting sphere with uniform volume free charge density $\rho_{\!f}$, both $W_E$ and $W_D$ energies are equal. The electric potentials of this test case are 
\begin{equation}
\label{eqVrho}
V_\textrm{in}(\vect{r}) = \frac{\rho_{\!f}\left((2\varepsilon_r+1)R^2-r^2\right)}{6\varepsilon_0\varepsilon_r},
\qquad 
V_\textrm{out}(\vect{r}) = \frac{\rho_{\!f} R^3}{3\varepsilon_0 r},
\end{equation}
and the electric field expressions are~\cite{GRIFFITHS2024} 
\begin{equation}
\label{eqErho}
\vect{E}_\textrm{in}(\vect{r})=\frac{\rho_{\!f}}{3\varepsilon_0\varepsilon_r}\vect{r},
\qquad
\vect{E}_\textrm{out}(\vect{r})=\frac{\rho_{\!f} R^3}{3\varepsilon_0 r^3}\vect{r}.
\end{equation}
The electrostatic energy of this sphere can be calculated with either of the integrals shown in row $\rho_{\!f}$ of Table~\ref{tab_es} (open space or finite space), the result of which is given in the rightmost column. This energy result is also found in~\cite{GRIFFITHS2024}.

\subsection{Surface charge density $\sigma_{\!f}$ or potential $V_0$}

The next two test cases appearing in Table~\ref{tab_es} have imposed electric potential (conducting sphere) or surface free charge density (conducting or not). The potentials are
\begin{equation}
V_\textrm{in}(\vect{r}) =V_0 = \frac{\sigma_{\!f} R}{\varepsilon_0},\qquad 
V_\textrm{out}(\vect{r}) =\frac{V_0R}{r} = \frac{\sigma_{\!f} R^2}{\varepsilon_0 r},
\end{equation}
and the electric field expressions are~\cite{GRIFFITHS2024,DECASTRO1991} 
\begin{equation}
\vect{E}_\textrm{in}(\vect{r})=\vect{0},
\qquad
\vect{E}_\textrm{out}(\vect{r})=\frac{\sigma_{\!f} R^2}{\varepsilon_0 r^3}\vect{r}.
\end{equation}

Specifying the total free charge $Q_{\!f}$ of the sphere (whether it is uniformly distributed in the volume or on the surface of the sphere) would also be possible, though this is not shown explicitly as it follows naturally from the equalities
\begin{equation}
Q_{\!f}=\sigma_{\!f} 4\pi R^2,\qquad Q_{\!f}=\rho_{\!f} 4\pi R^3/3.
\end{equation}

The relative permittivity of the sphere, when nonconducting, does not come into play given that the electric field inside the sphere is equal to zero. This is indicated with the mnemonic ``$\varepsilon_r$-ind.'' in row $\sigma_{\!f}$ of Table~\ref{tab_es} (meaning $\varepsilon_r$-independent). When the sphere is conducting, its exact conductivity value is not relevant since the charges are considered to be in equilibrium. 
With the above constraints, there is no bound charge and the free charge density~$\sigma_{\!f}$ is equal to the total charge density~$\sigma$ ($\sigma_{\!f}$ will be used for consistency). The resulting energy expression appears in the table.

\subsection{Volume charge density or potential with two spheres ($\rho_{\!f}\,;\,-\rho_{\!f}$ or $V_0\,;\,-V_0$)}

The first test case involving two objects consists of oppositely charged non-conducting spheres of densities $\rho_{\!f}$ and $-\rho_{\!f}$, both with $\varepsilon_r=1$. 
Since the spheres cannot be cross polarized, the total electric potential or field is the sum of the potentials or fields of each sphere, as given in (\ref{eqVrho}) and (\ref{eqErho}), when accounting for the center position of each sphere (namely $\vect{r}_{c1}$ and $\vect{r}_{c2}$). The total electrostatic energy contained in this configuration is the sum of two self energy terms and an interaction energy. The self energy of each sphere corresponds to that given above for the $\rho_{\!f}$ test case with the substitution $\varepsilon_r=1$. 
The interaction energy is calculated with the finite space integral 
\begin{equation}
W_{12}=\rho_{\!f1}\int_{\Omega_{s1}}  V_{\textrm{out}2}(\vect{r}-\vect{r}_{c2})\,d\Omega_{s1},
\end{equation}
where $\rho_{\!f1}$ was factored out of the integral and $V_{\textrm{out}2}$ represents the potential of the second sphere outside of its boundary. Since the integral is taken over the first sphere, the mean value property~\cite{HU2000} of harmonic functions is used to simplify the integral. This integral becomes equal to the volume of the sphere~1 multiplied by the potential of sphere~2 at the center of sphere~1, namely $V_{\textrm{out}2}(\vect{d}=\vect{r}_{c1}-\vect{r}_{c2})$. For the second sphere, $V_{\textrm{out}2}$ corresponds to the $V_\textrm{out}$ expression in (\ref{eqVrho}) but with a negative charge density. The resulting total energy appears in the rightmost column of the ``$\rho_{\!f}\,;\,-\rho_{\!f}$'' test case of Table~\ref{tab_es}. 

The final electrostatic energy test case consists of two conducting spheres with potentials $V_0$ and $-V_0$. With the method of images, this is known to be equivalent to a single sphere above a ground plane. The potential and electric field of this configuration involve Legendre polynomials and bi-spherical coordinates~\cite{MORSE1953}. These expressions are not well-suited for error evaluations in FEM software (as required for an L2 error norm for example). 

The electrostatic energy obtained from the charge appearing in reference~\cite{MORSE1953} is the simplest metric for benchmarking purposes. Alternate energy results also exist~\cite{BANERJEE2021}, where the radii of the two spheres need not be equal, but are not considered here. 
The resulting energy expression is given in the rightmost column of the ``$V_0\,;\,-V_0$'' test case of Table~\ref{tab_es}, where auxiliary variables such as $k$ and $\beta$ are introduced for conciseness. 
Although an infinite sum appears in this energy calculation, the $w(i)$ argument converges relatively quickly as~$i$ increases, the slowest convergence occurring when the spheres are very close to each other. For example, with a gap of $R/2$ between the spheres ($d=2.5R$ and $k=1.25$), it can be observed that $\sum_{i=0}^{10} w(i)\approx 1.19$ and an upper bound on the residual is on the order of $3.4\cdot10^{-7}$ (as obtained using an exact integral).

Test cases involving two polarized spheres are not listed, but can be obtained by applying the appropriate substitutions in the expressions of the magnetized spheres, as explained further in Section~\ref{ssec_mp}.

\section{Magnetostatic energy test cases}
\label{sec_mag}

For a linearly magnetizable material with a permanent magnetization $\vect{M}_{p}$ and a permeability $\mu=\mu_0\mu_r$, the magnetization is~\cite{BERE2001}
\begin{equation}
\label{eqMtot}
\vect{M}=(\mu_r-1)\vect{H}+\vect{M}_{p}
\end{equation}
and the constitutive relation is~\cite{PIRIOU2024,CAMPBELL2000}
\begin{equation}
\label{eqCRms}
\vect{B}=\mu_0(\vect{H}+\vect{M})=\mu\vect{H}+\mu_0\vect{M}_{p}
\end{equation}
where $\vect{B}$ is the magnetic field, $\vect{H}$ is the auxiliary field~\cite{GRIFFITHS2024} and $\mu_0\vect{M}_{p}$ represents the remanent magnetic field in the magnet (called $\vect{B}_r$). Linear permanent magnets are sometimes specified by their coercive $\vect{H}$ field instead (called $\vect{H}_c$). It is assumed that the permanent magnetization $\vect{M}_{p}$ of the linear magnet is given, and that the equalities below can be invoked as needed:
\begin{equation}
\label{eqMBrHc}
\mu_0\vect{M}_{p}=\vect{B}_r=\mu\vect{H}_c.
\end{equation}

Different approaches to energy calculations exist in magnetostatics~\cite{FU2004}. The $W_H$ integrals in Table~\ref{tab_ms} calculate the energy, while the $W_B$ integrals calculate the coenergy~\cite{BLUNDELL2001,LOVATT1999}. This is how energy is reported in FEMM~\cite{FEMM_PM} and is the chosen approach in the present work. It aligns with force calculations based on the virtual work method
\begin{equation}
\label{eqVwork}
F=-\frac{\delta W}{\delta x}=\frac{\delta W^\prime}{\delta x},
\end{equation}
where $W$ and $W^\prime$ are the energy and coenergy respectively, and a positive magnetic force~$F$ acts in the direction of increasing~$x$, consistent with the fact that the total magnetic energy should increase when separating attracting permanent magnets. 
Other authors follow a reversed terminology, where $W_H$ is the coenergy, and $W_B$ is the energy~\cite{CAMPBELL2000,BOSSAVIT1998}. In a linear material without permanent magnetization, energy and coenergy are equal.

\begin{table*}
\centering
\caption{Magnetostatic energy test cases for a single sphere (first three sources) and for two spheres (last three sources).\label{tab_ms}}
\renewcommand{\arraystretch}{1.5}
\begin{tabular*}{0.93\linewidth}{@{\extracolsep{\fill}} llllll@{}}

\toprule
Source & Material & Energy & Open Space & Finite Space & Result for Sphere(s) \\ \midrule

\multirow{2}{*}{${M}_p$} &
\multirow{2}{*}{Cond. or NC} & 
$W_H$ & 
$\frac{1}{2}\int_\Omega\mu H^2\,d\Omega$ &   
$-\frac{\mu_0}{2}\int_{\Omega_s} \vect{M}_p\cdot\vect{H}\,d\Omega_s$ & 
$\frac{2\pi\mu_0}{3(\mu_r+2)} R^3 M_{\!p}^2$ \\

& 
&
$W_B$& 
$\frac{1}{2}\int_\Omega\frac{1}{\mu} B^2\,d\Omega$ &
$ \frac{\mu_0}{2}\int_{\Omega_s}\frac{1}{\mu} \vect{M}_p\cdot\vect{B}\,d\Omega_s$ & 
$\frac{4\pi\mu_0}{3\mu_r(\mu_r+2)} R^3 M_{\!p}^2$\\[2pt]
\hline 

${J}_{\!f}$ & 
\multirow{2}{*}{NC; $\varepsilon_r=\mu_r=1$} & 
\multirow{2}{*}{$W_H=W_B$} & 
\multirow{2}{*}{$ \frac{1}{2}\int_\Omega \vect{H}\cdot\vect{B}\,d\Omega$} &
$\frac{1}{2}\int_{\Omega_s} \vect{J}_{\!f}\cdot\vect{A}\,d\Omega_s$ & 
$\frac{8\pi\mu_0}{315} R^7 \rho_{\!f}^2\omega^2$\\[1pt]
\cline{1-1}\cline{5-6}

${K}_{\!f}$ & 
& 
 &
 &
$\frac{1}{2}\int_{\Gamma_s} \vect{K}_{\!f}\cdot\vect{A}\,d\Gamma_s$ &
$\frac{4\pi\mu_0}{9} R^5 \sigma_{\!f}^2\omega^2$\\[1pt]
\hline

${M}_{\!p}\rightarrow\,;\,{M}_{\!p}\,+\!\leftrightarrow\!-$ &
& 
& 
&   
& 
$\frac{4\pi\mu_0}{9} \Bigl(1\pm\frac{2R^3}{d^3}\Bigr)R^3 M_{\!p}^2$ \\[1pt]
\cline{1-1}\cline{6-6}

${M}_{\!p}\uparrow\,;\,{M}_{\!p}\updownarrow\!\pm$ &
C or NC; $\mu_r=1$ & 
$W_H$ & 
$\frac{1}{2}\int_\Omega\mu H^2\,d\Omega$ &   
$-\frac{\mu_0}{2}\int_{\Omega_{s1}\cup\Omega_{s2}} \vect{M}_p\cdot\vect{H}\,d\Omega_s$ & 
$\frac{4\pi\mu_0}{9} \Bigl(1\pm\frac{R^3}{d^3}\Bigr)R^3 M_{\!p}^2$ \\[1pt]
\cline{1-1}\cline{6-6}

${M}_{\!p}\updownarrow\,;\,{M}_{\!p}\leftrightarrow$ &
& 
& 
&   
& 
$\frac{4\pi\mu_0}{9} R^3 M_{\!p}^2$ \\

\bottomrule

\end{tabular*}
\end{table*}

\subsection{Permanent magnetization $M_{p}$}
\label{ssec_mp}

The magnetic vector potentials inside and outside of the magnetized sphere are also calculated by solving the Laplace equation by separation of variables, making use of Legendre polynomials. An axisymmetric solution for the $\phi$ component ($A_\phi$) is applicable. This solution can be derived similarly to the one described in Section~\ref{ssec_pp}, by replacing $V$ with $A_\phi$ in~(\ref{eqBCes1}) and~(\ref{eqBCes3}), and by using the following equation in~(\ref{eqBCes2}):
\begin{equation}
\label{eqBCem2}
-\frac{1}{r\mu_0\mu_r}\frac{\partial (r A_{\phi\textrm{in}})}{\partial r} +\frac{M_{\!p}}{\mu_r}\sin(\theta) = -\frac{1}{r\mu_0}\frac{\partial (r A_{\phi\textrm{out}})}{\partial r} 
\end{equation}
The magnetization is directed in the positive $z$-axis ($\theta=0$). At the surface of the sphere, the boundary condition in~(\ref{eqBCem2}) corresponds to $H_{\parallel\textrm{in}}=H_{\parallel\textrm{out}}$, along with the use of~(\ref{eqCRms}). Equation~(\ref{eqCRms}) reduces to $\vect{B}=\mu_0\vect{H}$ outside of the sphere. 
Once solved, the vector potentials can be used to calculate the $\vect{B}$ fields in both regions with $\vect{B}=\nab\times\vect{A}$ (using spherical coordinates). When generalized for an arbitrary magnetization direction and expressed in a coordinate-independent form,
the vector potentials are
\begin{equation}
\label{eqAMinout}
\vect{A}_\textrm{in}(\vect{r}) = \frac{\mu_0}{\mu_r+2}\,\vect{M}_{p}\!\times \vect{r},\qquad
\vect{A}_\textrm{out}(\vect{r}) = \frac{\mu_0 R^3}{(\mu_r+2)r^3}\,\vect{M}_{p}\!\times \vect{r}.
\end{equation}
and magnetic fields are
\begin{equation}
\label{eqBM}
\begin{aligned}
&\vect{B}_\textrm{in}(\vect{r}) = \frac{2\mu_0}{\mu_r+2}\,\vect{M}_{p},\\
&\vect{B}_\textrm{out}(\vect{r}) = \frac{\mu_0 R^3}{(\mu_r+2)r^3}\left(\frac{3}{r^2}\left(\vect{M}_{p}\cdot\vect{r}\right)\vect{r}-\vect{M}_{p}\right).
\end{aligned}
\end{equation}
The $\vect{H}$ field follows naturally from~(\ref{eqCRms}). An uncentered sphere can be evaluated with $\vect{A}(\vect{r}-\vect{r}_c)$ and $\vect{B}(\vect{r}-\vect{r}_c)$. 
The fields obtained are consistent with implementations in ANSYS Maxwell~\cite{ANSYS} and FEMM~\cite{FEMM}, but differ from those in~\cite{BANOS1952} by a factor of $\mu_r$, where a different definition of permanent magnetization was used. 

The energy calculations follow using the appropriate field integrals in row ${M}_{\!p}$ of Table~\ref{tab_ms}, invoking either the open space or the finite space forms. 
The energy results for the sphere are also shown in the same row of the table. 
The $W_H$ expression generalizes the one reported in~\cite{BLUNDELL2001}, which is only valid for $\mu_r=1$. 

Since there are no free currents in the ${M}_{\!p}$ test case, it is also possible to solve the Laplace equation for a magnetic scalar potential~$\varphi$. This essentially represents a mapping from a magnetostatic problem to an electrostatic one (with a fictitious magnetic charge). 
As a verification step, it can be shown that the results in row ${P}_{\!p}$ of Table~\ref{tab_es} are identical to those in row ${M}_{\!p}$ of Table~\ref{tab_ms} when the following substitutions are applied~\cite{GRIFFITHS2024} 
\begin{equation}
\vect{D}\rightarrow\vect{B},\qquad
\vect{E}\rightarrow\vect{H},\qquad
\vect{P}_{\!p}\rightarrow\mu_0\vect{M}_{p},\qquad
\varepsilon_0\rightarrow\mu_0,
\end{equation}
to which $\varepsilon_r\rightarrow\mu_r$ (and thus $\varepsilon\rightarrow\mu$) can be added. Applying these substitutions to the electric potentials in~(\ref{eqVP}) also provides a convenient way to obtain the magnetic scalar potentials~$\varphi$ of the magnetized sphere. 
The substitutions apply only when there is no free charge or free current involved.

\subsection{Volume current density $J_{\!f}$}


The field source in the ${J}_{\!f}$ test case is a spinning free charge, uniformly distributed throughout the volume of the sphere. Since a free current is produced, the magnetic vector potential formulation is required. 
The sphere is assumed to be nonconducting, to avoid any charge displacement due to the Hall effect. Spinning bound charges and currents are also avoided by neglecting any polarization or magnetization inside the sphere ($\varepsilon_r=\mu_r=1$). In such a case, the free charge density~$\rho_{\!f}$ is equal to the total charge density~$\rho$ (though $\rho_{\!f}$ will be used for consistency). Although solid materials with $\mu_r\approx1$ are common, this is not the case for $\varepsilon_r$; nonetheless, it will be assumed that the solid sphere is composed of a porous foam with relative permittivity near~1.

The magnetic vector potential for the volumetrically charged spinning sphere can be derived from the spinning thin shell, whose potentials appear further below in the ${K}_{\!f}$ test case (see~(\ref{eqAk})). With the proper volume integrals, in which the surface density $\sigma_{\!f}$ is replaced by $\rho_{\!f}dr$, the magnetic vector potentials of the full sphere are found to be
\begin{equation}
\label{eqAj}
\vect{A}_\textrm{in}(\vect{r}) = \frac{\mu_0\rho_{\!f}(5R^2-3r^2)}{30}\,\boldsymbol{\omega}\times\vect{r},\qquad
\vect{A}_\textrm{out}(\vect{r}) =  \frac{\mu_0\rho_{\!f} R^5}{15r^3}\,\boldsymbol{\omega}\times\vect{r},
\end{equation}
and the magnetic fields are
\begin{equation}
\begin{aligned}
&\vect{B}_\textrm{in}(\vect{r}) = \mu_0 \rho_{\!f}\left( 
\left(\frac{R^2}{3r^2}-\frac{1}{5}\right)\left(\boldsymbol{\omega}\cdot\vect{r}\right)\vect{r}
-\left(\frac{R^2}{3r^2}-\frac{2}{5}\right)\left(\boldsymbol{\omega}\times\vect{r}\right)\times\vect{r}\right),
\\
&\vect{B}_\textrm{out}(\vect{r}) = \frac{\mu_0 \rho_{\!f} R^5}{15r^3}\left(\frac{3}{r^2}\left(\boldsymbol{\omega}\cdot\vect{r}\right)\vect{r}-\boldsymbol{\omega}\right).
\end{aligned}
\end{equation}
As before, the magnetic energy is best obtained with a sphere centered at the origin and with integration performed in spherical coordinates. The resulting energy is given in the rightmost column of row ${J}_{\!f}$ in Table~\ref{tab_ms}. 
Although the charges are radially accelerated in this configuration, no electromagnetic radiation occurs and the fields are entirely static~\cite{HEALY2002}.

\subsection{Surface current density $K_{\!f}$}

The spinning solid sphere with a surface free charge density $\sigma_{\!f}$ implies the existence of a surface current. The sphere is also assumed to be non-conducting, to avoid any charge displacement due to the Hall effect, along with $\varepsilon_r=\mu_r=1$. 
The magnetic vector potentials of the sphere are equal to those of the spinning charged thin shell, which are~\cite{GRIFFITHS2024}: 
\begin{equation}
\label{eqAk}
\vect{A}_\textrm{in}(\vect{r}) = \frac{\mu_0\sigma_{\!f} R}{3}\,\boldsymbol{\omega}\times\vect{r},\qquad
\vect{A}_\textrm{out}(\vect{r}) = \frac{\mu_0\sigma_{\!f}R^4}{3r^3}\,\boldsymbol{\omega}\times\vect{r}.
\end{equation}

Since the spinning shell produces the same fields as a uniformly magnetized sphere having the reference permeability~\cite{GRIFFITHS2024}, 
the magnetic fields are~\cite{DECASTRO1991}
\begin{equation}
\vect{B}_\textrm{in}(\vect{r}) = \frac{2\mu_0 \sigma_{\!f}R}{3}\,\boldsymbol{\omega},\qquad
\vect{B}_\textrm{out}(\vect{r}) = \frac{\mu_0 \sigma_{\!f} R^4}{3r^3}\left(\frac{3}{r^2}\left(\boldsymbol{\omega}\cdot\vect{r}\right)\vect{r}-\boldsymbol{\omega}\right).
\end{equation}
The energy can be obtained by integration, the result of which is shown in row ${K}_{\!f}$ of Table~\ref{tab_ms}. 
Substituting $\vect{M}_p=\sigma_{\!f}R\,\boldsymbol{\omega}$ and  $\mu_r=1$ into the $W_B$ energy of the ${M}_{\!p}$ test case also yields the energy in the ${K}_{\!f}$ test case.

\subsection{Permanent magnetization with two spheres ($M_{\!p\,};M_{\!p}$)}

When two magnetized spheres with $\mu_r=1$ are used as the test objects, the total magnetic energy depends on both magnetization directions. Only the $W_H$ energy is used here. 
The potentials and fields superimpose given the absence of cross magnetization. The total magnetic energy is the sum of two self energy terms and an interaction energy. 
The self energy of a magnetized sphere corresponds to the first result given in Table~\ref{tab_ms}, with the substitution $\mu_r=1$. 

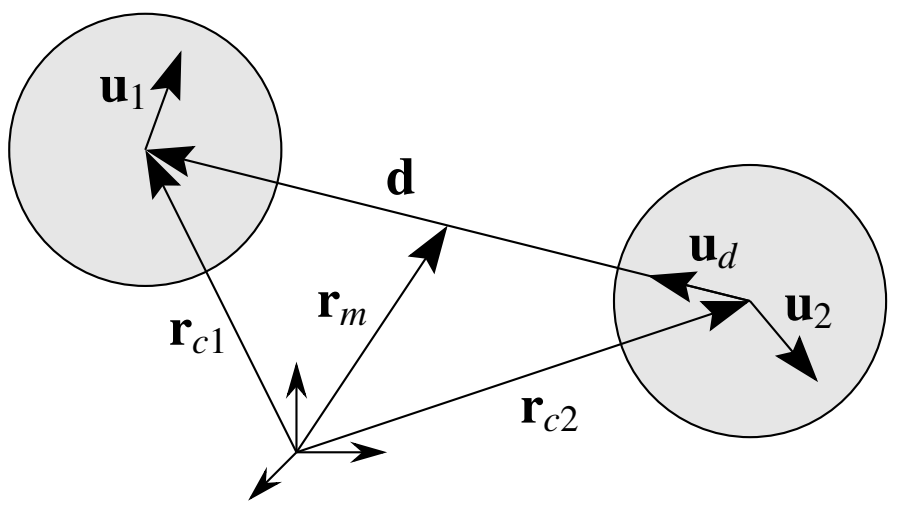
\begin{figure}
\centerline{

\begin{tikzpicture}
\def\scl{1.0}

\def\rs{0.9}
\def\un{0.7}
\coordinate (O) at (0,0);
\coordinate (C1) at (-1,2);
\coordinate (C2) at (3,1);

\draw[fill=gray!20!white] (C1) circle (\rs);
\draw[fill=gray!20!white] (C2) circle (\rs);

\draw [arrows = {-Stealth[inset=1.5pt,width=6pt, length=9pt]}] (O) to (C1); 
\draw[] ([shift=({130:1})]O) node[] {\scalebox{\scl}{$\vect{r}_{c1}$}};

\draw [arrows = {-Stealth[inset=1.5pt,width=6pt, length=9pt]}] (O) to (C2); 
\draw[] ([shift=({8.5:1.7})]O) node[] {\scalebox{\scl}{$\vect{r}_{c2}$}};

\draw [arrows = {-Stealth[inset=1.5pt,width=6pt, length=9pt]}] (C2) to (C1); 
\draw[] ([shift=({-6:1.7})]C1) node[] {\scalebox{\scl}{$\vect{d}$}};

\draw [arrows = {-Stealth[inset=1.5pt,width=6pt, length=9pt]}] (O) to ([shift=({165.964:2.06155})]C2); 
\draw[] ([shift=({72:1})]O) node[] {\scalebox{\scl}{$\vect{r}_m$}};

\draw [arrows = {-Stealth[width=4pt, length=7pt]}] (O) to ++(0,0.6); 
\draw [arrows = {-Stealth[width=4pt, length=7pt]}] (O) to ++(0.6,0); 
\draw [arrows = {-Stealth[width=4pt, length=7pt]}] (O) to ++(-0.32,-0.32); 

\draw [arrows = {-Stealth[inset=1.5pt,width=6pt, length=9pt]}]  (C1) to ([shift=({70:\un})]C1); 
\draw[] ([shift=({110:0.4})]C1) node[] {\scalebox{\scl}{$\vect{u}_1$}};

\draw [arrows = {-Stealth[inset=1.5pt,width=6pt, length=9pt]}]  (C2) to ([shift=({-50:\un})]C2);
\draw[] ([shift=({-10:0.4})]C2) node[] {\scalebox{\scl}{$\vect{u}_2$}};

\draw [arrows = {-Stealth[inset=1.5pt,width=6pt, length=9pt]}]  (C2) to ([shift=({165.964:\un})]C2);
\draw[] ([shift=({126:0.4})]C2) node[] {\scalebox{\scl}{$\vect{u}_d$}};

\end{tikzpicture}

}
\caption{Vectors for two spheres.\label{figVect}}
\end{figure}

The interaction energy between two point dipoles is~\cite{EDWARDS2017}
\begin{equation}
\label{eqW12}
W_{12}=\frac{\mu_0}{4\pi}\left(\frac{\vect{m}_1\cdot \vect{m}_2}{d^3} - 3\frac{(\vect{m}_1\cdot \vect{d})(\vect{m}_2\cdot \vect{d})}{d^5}\right),
\end{equation}
where $\vect{m}_1$ and $\vect{m}_2$ are the two dipole moments, and $\vect{d}$ is the position vector from dipole~2 to dipole~1 (reversing this vector has no effect on the energy in this case). The correspondence between (\ref{eqW12}) and the interaction energy of two spheres is possible using the mean value property of harmonic functions, along with
\begin{equation}
\label{eqm1m2}
\vect{m}_1 = \frac{4}{3}\pi R^3 M_{\!p}\, \vect{u}_1,\qquad \vect{m}_2 = \frac{4}{3}\pi R^3 M_{\!p}\, \vect{u}_2,
\end{equation}
where $\vect{u}_1$ and $\vect{u}_2$ are the unit vectors of the magnetization directions of each sphere. Adding the two self energies of the spheres to the interaction energy, the total energy becomes
\begin{equation}
\label{eqWMM}
W_{H}=\frac{4\pi\mu_0}{9}\left(1+\big(\vect{u}_1\cdot \vect{u}_2 - 3(\vect{u}_1\cdot \vect{u}_d)(\vect{u}_2\cdot \vect{u}_d)\big)\frac{R^3}{d^3}\right)R^3M_{\!p}^2,
\end{equation}
where $\vect{d}=\vect{r}_{c1}-\vect{r}_{c2}$ and $\vect{u}_d$ is the unit vector of $\vect{d}$, as shown in Fig.~\ref{figVect}. Replacing the symbols ``$1+$'' in (\ref{eqWMM}) by ``$2-$'' yields the expression for the coenergy ($W_B$).

Three particular cases for the energy of two magnetized spheres are presented in Table~\ref{tab_ms}, using the following notation:
\begin{itemize}
\item ${M}_{\!p}\rightarrow\,;\,{M}_{\!p}\,+\!\leftrightarrow\!-$~: direction $\vect{d}$ and magnetization directions $\vect{u}_1$ and $\vect{u}_2$ are all collinear; the ``$-$'' sign is used when $\vect{u}_1=\vect{u}_2$ and the ``$+$'' sign is used when $\vect{u}_1=-\vect{u}_2$;
\item ${M}_{\!p}\uparrow\,;\,{M}_{\!p}\updownarrow\!\pm$~: magnetization directions $\vect{u}_1$ and $\vect{u}_2$ are parallel to each other and orthogonal to the direction vector $\vect{d}$; the ``$-$'' sign is used when $\vect{u}_1=-\vect{u}_2$ and the ``$+$'' sign is used when $\vect{u}_1=\vect{u}_2$;
\item ${M}_{\!p}\updownarrow\,;\,{M}_{\!p}\leftrightarrow$~: both magnetization directions $\vect{u}_1$ and $\vect{u}_2$ are orthogonal and one is parallel to $\vect{d}$.
\end{itemize}
These serve as quick and practical test cases for convenience (with simpler energy formulae), whereas (\ref{eqWMM}) is the more general result. Equivalent test cases for two permanently polarized spheres can also be obtained by applying the substitutions given at the end of Section~\ref{ssec_mp}.

\section{Force and torque test cases for two spheres}
\label{sec_f2s}

The previous test cases comprising two spheres also involve force, and in one particular instance, torque. Electromagnetic force and torque are often obtained using the virtual work method or the Maxwell stress tensor~\cite{GRIFFITHS2024,BERMUDEZ2017}. This tensor is denoted $\overleftrightarrow{\vect{T}}$ and contains both electric and magnetic fields; $\overleftrightarrow{\vect{T}}_E$ is this tensor with only the electric field, and $\overleftrightarrow{\vect{T}}_B$ is the tensor with only the magnetic field. 

\begin{table*}
\centering
\caption{Force and torque test cases for two spheres ($\vect{r}_m$ is the midpoint between the spheres).\label{tab_2s}}
\renewcommand{\arraystretch}{1.5}
\begin{tabular*}{0.88\linewidth}{@{\extracolsep{\fill}} lllll@{}}

\toprule
Source & Material & Quantity & Finite space & Magnitude on one sphere \\ \midrule

$\rho_{\!f}\,;\,-\rho_{\!f}$ &
NC; $\varepsilon_r=1$ & 
\multirow{4}{*}{$\vect{F}$} & 
\multirow{4}{*}{$\int_{\Gamma_{s}} \overleftrightarrow{\vect{T}}_E\cdot \vect{n}\,d\Gamma_{s}$} & 
$\frac{4\pi}{9\varepsilon_0d^2} R^6\rho_{\!f}^2 $ \\[2pt]
\cline{1-2}\cline{5-5}

&
& 
&
&
$4\pi\varepsilon_0 V_0^2 \sum_{i=0}^\infty f(i)\,$,~with\\

$V_0\,;\,-V_0$ &
Conducting & 
&
&
$f(i)=\frac{(2i+1)\gamma(i)}{(\gamma(i)-1)^2}-\frac{k}{(\gamma(i)-1)\beta},~k=\frac{d}{2R}$,\\

&
& 
&
&
$\gamma(i)=(k+\beta)^{2i+1},~\beta=\sqrt{k^2-1}$\\
\hline

${M}_{\!p}\leftrightarrow\,;\,{M}_{\!p}\leftrightarrow$ &
\multirow{4}{*}{C or NC; $\mu_r=1$} & 
& 
&   
$\frac{8\pi\mu_0}{3d^4} R^6 M_{\!p}^2$ \\[2pt]
\cline{1-1}\cline{5-5}

${M}_{\!p}\updownarrow\,;\,{M}_{\!p}\updownarrow$ &
& 
$\vect{F}$ &   
$\int_{\Gamma_{s}} \overleftrightarrow{\vect{T}}_B\cdot \vect{n}\,d\Gamma_{s}$ & 
\multirow{2}{*}{$\frac{4\pi\mu_0}{3d^4} R^6 M_{\!p}^2$} \\
\cline{1-1}

${M}_{\!p}\updownarrow\,;\,{M}_{\!p}\leftrightarrow$ &
& 
& 
&   
 \\[2pt]
 \cline{1-1}\cline{3-5}
 
${M}_{\!p}\updownarrow\,;\,{M}_{\!p}\leftrightarrow$ &
& 
$\boldsymbol{\tau}$ & 
$\int_{\Gamma_{s}} (\vect{r}-\vect{r}_m)\times(\overleftrightarrow{\vect{T}}_B\cdot \vect{n})\,d\Gamma_{s}$ &   
$\frac{2\pi\mu_0}{9d^3} R^6 M_{\!p}^2$ \\[1pt]

\bottomrule 

\end{tabular*}
\end{table*}

In the tensor approach, volume integrals involving the divergence of $\overleftrightarrow{\vect{T}}$ can be used to obtain force and torque. 
To reduce the order of derivation required in the numerical evaluation, the divergence theorem is often used to express these as surface integrals instead. The integrals are given in the fourth column in Table~\ref{tab_2s}, where $\vect{n}$ is the unit vector normal to the surface of integration (pointing outward). A more practical computation of these surface integrals is actually performed using a thin probe-layer of elements in FEM simulations, as explained below in Section~\ref{ssec_fts}.

\subsection{Electrostatic force}

For two conducting spheres or two uniformly charged nonconducting spheres, the electrostatic force is more readily obtained analytically using the virtual work method (rather than the tensor approach). Since the force and direction vectors are collinear, the magnitude of the force on a given sphere is
\begin{equation}
F_{(Q)}=\frac{\delta W_E}{\delta d}\quad(\text{constant~}Q),\qquad F_{(V)}=-\frac{\delta W_E}{\delta d}\quad(\text{constant~}V).
\end{equation}
The two derivatives above are used in the ``$\rho_{\!f}\,;\,-\rho_{\!f}$'' and ``$V_0\,;\,-V_0$'' test cases respectively, with the corresponding energy expressions from Table~\ref{tab_es}. The resulting force magnitudes for the two cases are given in the first two rows of the rightmost column in Table~\ref{tab_2s}. The force vectors are $\vect{F}_1=-F\,\vect{u}_d$ and $\vect{F}_2=-\vect{F}_1$.

\subsection{Magnetostatic force}

The magnetostatic force between two uniformly magnetized spheres can be obtained from the force between two magnetic dipoles~\cite{EDWARDS2017} and the equalities in~(\ref{eqm1m2}), resulting in the forces
\begin{equation}
\label{eqFM}
\begin{aligned}
\vect{F}_1 = &\frac{4\pi\mu_0}{3d^4}\Big( 
(\vect{u}_1\cdot \vect{u}_2)\vect{u}_d+
(\vect{u}_d\cdot \vect{u}_2)\vect{u}_1+
(\vect{u}_d\cdot \vect{u}_1)\vect{u}_2\\
&-5(\vect{u}_d\cdot \vect{u}_1)(\vect{u}_d\cdot \vect{u}_2)\vect{u}_d
 \Big)\,R^6 M_{\!p}^2,\qquad \vect{F}_2=-\vect{F}_1.
 \end{aligned}
\end{equation}
Table~\ref{tab_2s} presents three particular test cases for the force, with magnetization indicators similar to those used in the magnetic energy instances. The magnetization directions correspond to the cases:
\begin{itemize}
\item ${M}_{\!p}\leftrightarrow\,;\,{M}_{\!p}\leftrightarrow$~: all three unit vectors are collinear;
\item ${M}_{\!p}\updownarrow\,;\,{M}_{\!p}\updownarrow$~: magnetizations are parallel, both of which are orthogonal to the direction vector;
\item ${M}_{\!p}\updownarrow\,;\,{M}_{\!p}\leftrightarrow$~: only one magnetization direction is collinear with $\vect{d}$, the other is orthogonal to $\vect{d}$.
\end{itemize}

The torque on a given sphere has two contributions~\cite{EDWARDS2017}. The first is a local dipole torque and the second arises from the lever arm of the force. For sphere~1, the local dipole torque is
\begin{equation}
\boldsymbol{\tau}_{1A}=
\,\vect{M}_{p1}\!\times\int_{\Omega_{s1}} \! \vect{B}_{\textrm{out}2}(\vect{r}-\vect{r}_{c2})\,d\Omega_{s1}.
\end{equation}
where $\vect{M}_{p1}$ was factored out of the integral and $\vect{B}_{\textrm{out}2}$ represents the field of the second sphere outside of its boundary. Since the integral is taken over sphere~1, the mean value property of harmonic functions is used to simplify the integral. This integral becomes equal to the volume of sphere~1 multiplied by the field of sphere~2 at the center of sphere~1, namely $\vect{B}_{\textrm{out}2}(\vect{d}=\vect{r}_{c1}-\vect{r}_{c2})$. For the second sphere, $\vect{B}_{\textrm{out}2}$ corresponds to $\vect{B}_\textrm{out}$ given in (\ref{eqBM}) with $\mu_r=1$ and $\vect{M}_{p}=\vect{M}_{p2}$. The resulting torque becomes
\begin{equation}
\boldsymbol{\tau}_{1A}=\frac{4\pi R^3}{3}\,\vect{M}_{p1}\!\times 
\left(\frac{\mu_0 R^3}{3d^3}\left(\frac{3}{d^2}\left(\vect{M}_{p2}\cdot\vect{d}\right)\vect{d}-\vect{M}_{p2}\right)\right).
\end{equation}
Since both spheres have the same magnitude of dipole moment per unit volume, this can be simplified to 
\begin{equation}
\boldsymbol{\tau}_{1A}=\frac{4\pi\mu_0 R^6 M_{\!p}^2}{9d^3}\,\vect{u}_1\times \big(3\left(\vect{u}_2\cdot \vect{u}_d \right)\vect{u}_d-\vect{u}_2\big).
\end{equation}
The second contribution of torque on sphere~1 comes from the force $\vect{F}_1$ in~(\ref{eqFM}). The midpoint between the spheres (at $\vect{r}_m=\vect{r}_{c2}+\vect{d}/2$) is used as reference point for the lever arm. The second torque is
\begin{equation}
\boldsymbol{\tau}_{1B}=\frac{d}{2}\vect{u}_d\times{\vect{F}_1}=
\frac{2\pi\mu_0 R^6 M_{\!p}^2}{3d^3}\,\vect{u}_d\times 
\big((\vect{u}_d\cdot \vect{u}_2)\vect{u}_1+
(\vect{u}_d\cdot \vect{u}_1)\vect{u}_2\big).
\end{equation}
The total torque on sphere~1 is $\boldsymbol{\tau}_{1A}+\boldsymbol{\tau}_{1B}$ and the total torque on sphere~2 is equal in magnitude and opposite in direction, therefore
\begin{equation}
\begin{aligned}
\boldsymbol{\tau}_{1}=&\frac{\pi\mu_0 R^6 M_{\!p}^2}{9d^3}\Big(
4\,\vect{u}_1\times \big(3\left(\vect{u}_2\cdot \vect{u}_d \right)\vect{u}_d-\vect{u}_2\big) \\&+
6\,\vect{u}_d\times 
\big((\vect{u}_d\cdot \vect{u}_2)\vect{u}_1+
(\vect{u}_d\cdot \vect{u}_1)\vect{u}_2\big)
\Big),\qquad \boldsymbol{\tau}_{2}=-\boldsymbol{\tau}_{1}.
\end{aligned}
\end{equation}
Among the three particular magnetic-force cases mentioned above, only the ``$\updownarrow\,;\,\leftrightarrow$'' configuration has non-zero torque. 
The magnitude of the total torque about $\vect{r}_m$ on a given sphere in this situation is shown in the bottom-right in Table~\ref{tab_2s}.

\section{Electromagnetic momentum test cases for the single sphere\label{sec_em}}

None of the previous test cases intrinsically require a volume integral of a vector quantity. Energy integrals involve either the norm of a vector or the scalar product of two vectors, while force and torque integrals are typically only evaluated over a thin probe layer in the FEM. 
A vector quantity that can be integrated in all of space is electromagnetic momentum~\cite{GRIFFITHS2012}. 
The momentum density~$\vect{g}$ and angular momentum density~$\boldsymbol{\ell}$ strictly ascribed to the electromagnetic fields are~\cite{GRIFFITHS2024,MCDONALD2026}, 
\begin{equation} 
\vect{g}=\varepsilon_0\vect{E}\times\vect{B},\qquad
\boldsymbol{\ell}=\vect{r}\times\vect{g}.
\end{equation}
The linear momentum $\vect{p}$ and angular momentum $\vect{L}$ are the volume integrals of $\vect{g}$ and $\boldsymbol{\ell}$ respectively, and can be non-zero even in static fields. 
Each of the test cases proposed below contains only one form of electromagnetic momentum (either $\vect{p}$ or $\vect{L}$). The Coulomb gauge ($\nab\cdot\vect{A}=0$) is assumed in any finite space integral involving~$\vect{A}$.

\begin{table*}
\centering
\caption{Electromagnetic momentum test cases for the single sphere.\label{tab_em}}
\renewcommand{\arraystretch}{1.5}
\begin{tabular*}{0.9\linewidth}{@{\extracolsep{\fill}} llllll@{}}

\toprule
Source & Material & Momentum & Open Space & Finite Space & Result for Sphere  \\ \midrule

\multirow{2}{*}{${P}_{\!p}\,,\,{M}_{\!p}$} &
\multirow{2}{*}{NC} & 
\multirow{2}{*}{$\vect{p}$} & 
\multirow{2}{*}{$\varepsilon_0\int_\Omega \vect{E} \times \vect{B}\,d\Omega$} & 
$\varepsilon_0\mu_0\int_{\Omega_s} \vect{E} \times \vect{M}\,d\Omega_s$ & 
\multirow{2}{*}{$\frac{4\pi\mu_0}{(\mu_r+2)(\varepsilon_r+2)} R^3\vect{M}_p\!\times\vect{P}_{\!p}$}
\\

 & 
 & 
 & 
 & 
$=\frac{1}{2}\int_{\Omega_s} \vect{B} \times \vect{P}\,d\Omega_s$ & 
\\[2pt]
\hline

$\rho_{\!f}\,,\,{M}_{\!p}$ &
NC & 
& 
&
\multirow{2}{*}{$\int_{\Omega_s} \vect{r} \times (\rho\vect{A})\,d\Omega_s$} & 
$\frac{8\pi\mu_0(5\varepsilon_r-2)}{45\varepsilon_r(\mu_r+2)} R^5\rho_{\!f} \vect{M}_p$ \\[2pt]
\cline{1-2}\cline{6-6}

${J}_{\!f}$ & 
NC; $\varepsilon_r=\mu_r=1$ & 
 & 
 & 
 &
$\frac{16\pi\mu_0}{315} R^7\rho_{\!f}^2 \boldsymbol{\omega}$\\[1pt]
\cline{1-2}\cline{5-6}

$V_0\,,\,{M}_{\!p}$ &
C & 
$\vect{L}$ & 
$\varepsilon_0\int_\Omega \vect{r} \times (\vect{E} \times \vect{B})\,d\Omega$ &
 &  
$\frac{8\pi\varepsilon_0\mu_0}{3(\mu_r+2)} R^3V_0\vect{M}_p$
\\[1pt]
\cline{1-2}\cline{6-6}

$\sigma_{\!f}\,,\,{M}_{\!p}$ &
NC ($\varepsilon_r$-ind.) or C & 
 & 
 &
$\int_{\Gamma_s} \vect{r} \times (\sigma\vect{A})\,d\Gamma_s$ &
$\frac{8\pi\mu_0}{3(\mu_r+2)} R^4\sigma_{\!f} \vect{M}_p$
\\[1pt]
\cline{1-2}\cline{6-6}

${K}_{\!f}$ & 
NC; $\varepsilon_r=\mu_r=1$ & 
 & 
 &
 &
$\frac{8\pi\mu_0}{9} R^5\sigma_{\!f}^2 \boldsymbol{\omega}$
\\

\bottomrule

\end{tabular*}
\end{table*}

\subsection{Permanent polarization and magnetization}

The fields of the permanently polarized and magnetized sphere, labeled ``$\vect{P}_{\!p}\,,\,\vect{M}_p$'', carry linear electromagnetic momentum. The first row in Table~\ref{tab_em} shows three possible field integrals: one open space and two finite space integrals. The finite space integrals require the total polarization $\vect{P}$ or magnetization $\vect{M}$, as defined in~(\ref{eqPtot}) and~(\ref{eqMtot}). 
The resulting linear momentum is most easily obtained using a finite space integral. Since vectors $\vect{E}$, $\vect{P}_{\!p}$, $\vect{P}$, and $\vect{B}$ are uniform inside the sphere, and when invoking~(\ref{eqPtot}) for the total polarization $\vect{P}$, the second finite space integral shown in the table can be calculated as follows
\begin{equation}
\vect{p}=\frac{1}{2}\int_{\Omega_s} \vect{B} \times \vect{P}\,d\Omega_s=
\frac{2\pi}{3}R^3\vect{B}\times\left(\varepsilon_0(\varepsilon_r-1)\vect{E}+\vect{P}_{\!p}\right).
\end{equation}
Using~(\ref{eqEP}) and~(\ref{eqBM}), the momentum becomes equal to
\begin{equation}
\label{eqPMp}
\vect{p}=\frac{4\pi\mu_0}{(\mu_r+2)(\varepsilon_r+2)} R^3\,\vect{M}_p\!\times\vect{P}_{\!p}.
\end{equation}
The expression above generalizes those reported in~\cite{GRIFFITHS2024,BABSON2009}, 
which are valid only for $\varepsilon_r=\mu_r=1$. 

The linear electromagnetic momentum is canceled by a hidden momentum~\cite{BABSON2009,CORREA2020}, which is also present in this configuration, for a total linear momentum of zero.

\subsection{Volume and surface current densities}

The ${J}_{\!f}$ and ${K}_{\!f}$ test cases comprise only a single source (spinning charges), the fields of which carry electromagnetic angular momenta. These momenta are defined with respect to the center of the sphere and are most conveniently calculated with the finite space integrals listed in Table~\ref{tab_em}, using the vector potentials $\vect{A}_\textrm{in}$ in~(\ref{eqAj}) and~(\ref{eqAk}). The resulting momenta expressions are given in the rightmost column in rows ${J}_{\!f}$ and ${K}_{\!f}$ of the table. 
The angular momentum for the surface current case also appears in~\cite{DECASTRO1991}. 

Although the finite space integrals involve the total charge densities,  these are equal to the free charge densities here since there is no bound charge. This is because $\varepsilon_r=1$ and $\vect{P}_{\!p}=\vect{0}$. 
A verification can be done concerning the angular momenta, which must be consistent with their magnetostatic energy counterparts according to~\cite{GAUTHIER2006} 
\begin{equation}
W_B=\vect{L}\cdot\boldsymbol{\omega}/2.
\end{equation} 

In general, there is no equivalent hidden angular momentum to systematically cancel out the electromagnetic angular momentum~\cite{BABSON2009,HNIZDO1992}.

\subsection{Magnetization and electric charge}

The angular momenta for the three remaining test cases in Table~\ref{tab_em}, labeled ``$\rho_{\!f}\,,\,\vect{M}_p$'', ``$V_0\,,\,\vect{M}_p$'' and ``$\sigma_{\!f}\,,\,\vect{M}_p$'', are calculated by using the magnetic vector potentials of the magnetized sphere in~(\ref{eqAMinout}), along with the appropriate charge density or electric potential. 
With the exception of the ``$\rho_{\!f}\,,\,\vect{M}_p$'' case, the total charge densities $\rho$ and $\sigma$ appearing in the finite space integrals are equal to the free charge densities, given the constraints imposed. The momentum in the ``$\rho_{\!f}\,,\,\vect{M}_p$'' case is most conveniently calculated using the open space integral, since~$\rho$ is not necessarily equal to~$\rho_{\!f}$ in the finite space integral (and the presence of bound charges is more cumbersome).

\section{Numerical experiments}
\label{sec_sims}

A set of FEM simulation examples is provided to illustrate the application of the open boundary test cases. 
The free and open-source ONELAB software~\cite{ONELAB} (composed of Gmsh 4.15.2 and GetDP 3.5.0) is used in the numerical experiments. The simulation results are compared to the corresponding analytical expressions from the previous sections. 
The effects of the various parameters are not examined exhaustively in this work. 
Mesh convergence and error analysis are performed in Appendix~\ref{apx_err_conv} for a few configurations for illustrative purposes. Formulas involving $\vect{r}$ in~(\ref{eqJK}) are generalized for a sphere centered at an arbitrary position $\vect{r}_c$ by applying the substitution $\vect{r}\rightarrow(\vect{r}-\vect{r}_c)$. 

All simulations are performed with the geometrical and physical parameters given in Table~\ref{tab_params}. 
The permanent polarization is consistent with that of a linear electret material~\cite{SESSLER1987}, while the permanent magnetization is consistent with that of a linear ferrite magnet~\cite{PIRIOU2024}. 
Sphere offsets (displacements from the origin) are in the $z$-axis, and symmetry is exploited to reduce the simulation space to one half or one quarter of the full domain, depending on the case. 

\begin{table*}
\centering
\caption{Simulation parameters.\label{tab_params}}
\renewcommand{\arraystretch}{1.4}
\begin{tabular*}{0.78\linewidth}{@{\extracolsep{\fill}} ll@{}}

\toprule
Geometrical parameters & Physical parameters  \\ \midrule

Sphere radius: $R=1$~cm &
$\varepsilon_r=3.1$ (or 1 when required)
\\

Inner boundary radius: $r_b=5$~cm &
$\mu_r=1.6$ (or 1 when required)
\\

Exterior boundary radius: $r_e=10$~cm &
$P_{p}=5\cdot10^{-5}$~C/m$^2$
\\

3$^\textrm{rd}$ order IABC, layer thickness: $t=5/3$~cm &
$M_p=3.2\cdot10^{5}$~A/m; ($H_c=M_p/\mu_r=2\cdot10^{5}$~A/m)
\\

Sphere~1 offset from origin: $2$~cm &
$\rho_{\!f}=8\cdot10^{-5}$~C/m$^3$
\\

Sphere~2 opposite offset: $1$~cm ($d=3$~cm)&
$V_0=100$~V, and $\sigma_{\!f}=\varepsilon_0V_0/R$
\\

Probe layer interior radius: $1.5$~cm (sphere~1) &
$\omega=10^{3}$~rad/s
\\

\bottomrule

\end{tabular*}
\end{table*}

The first-order ABC method (called ABC-1) uses the $r=r_e$ radius as its outer boundary, such that the entire finite domain is used as an interior region. The truncation method is also applied at the $r=r_e$ radius. For the third-order IABC method (called IABC-3), the space within $r_b<r<r_e$ is allocated to the three layers. This same space is also used as the mapping destination of the $r>r_b$ space in the shell method, as shown in (\ref{eqshell}). The particular choice $r_e=2r_b$ provides a smoother transition at the inner boundary in the shell method, yielding $d\tilde{r}/dr=1$ at $r=r_b$. 

IABC layers do not normally need to be as thick. However, in the present approach, the interior region $r<r_b$ is designated as the minimum common space for all methods, and the remaining space in $r_b<r<r_e$ is used by each method as required for its boundary treatment. 
This makes it such that the mesh does not have to be regenerated each time the boundary method is changed in the numerical experiments. The IABC-3 coefficients are calculated with a SageMath script to support the case $\delta=1/3$ (which is not treated in~\cite{MEEKER2014}). 

All simulations are based on 3D models with certain symmetry restrictions for efficiency. 
Cases for which the centers of the spheres and the $\vect{P}_{\!p}$, $\vect{M}_p$ and $\boldsymbol{\omega}$ vectors are aligned with the axis of symmetry are suitable for axisymmetric simulations (not used here).
It is assumed that finite space integrals can make use of imposed parameters directly during post-processing. 
For example, in the~$V_0$ test case, $\sigma_{\!f}$ is assumed to be an unknown and should be obtained by post-processing means; however, in the $\sigma_{\!f}$ test case, it is not an unknown and can directly be used in a post-processing expression (it is the potential that must instead be evaluated in post-processing). 

The simulations use second order geometrical elements and interpolation functions over a static mesh. Simulations are not pushed to attain maximum precision, they are instead tuned to execute relatively quickly (on the order of a minute or two at most for the majority of the cases). The mesh sizes range from approximately 110k tetrahedra (quarter of domain, one sphere) to 340k tetrahedra (half of domain, two spheres). All simulation files and the SageMath IABC coefficient script are available online~\cite{SIMFILES}.

\subsection{Electrostatic and magnetostatic energy}
\label{ssec_emres}

Simulation examples for the electrostatic and magnetostatic energy test cases are presented in Tables~\ref{tab_ess} and~\ref{tab_mss}. The ABC-1 results are obtained with the finite space integrals, since the exterior region is not present with this method. For the other three boundary methods (truncation, IABC-3 and shell), open space and finite space integrals produce equal energy results to eight significant digits. 
Aside from the analytical (or reference) energy values, the quantities reported in the tables are relative differences with respect to the analytical values. These relative differences are expressed either in percentages or in parts per million (ppm), with 1~ppm equal to 0.0001$\%$. 

The magnetization directions for the experiments can be understood as follows: in a source configuration mnemonic, the $z$-axis points to the left and the $x$-axis points upward; for example, in the ``${M}_{\!p}\leftarrow\,;\,{M}_{\!p}\uparrow$'' test case (last line in Table~\ref{tab_mss}), sphere~1 is magnetized in the positive~$z$ direction, while sphere~2 is magnetized in the positive~$x$ direction.

With the equivalence shown at the end of Section~\ref{ssec_mp}, it could be expected that the relative differences for the ${P}_{\!p}$ and ${M}_{\!p}$ test cases would be identical (especially since both simulations used scalar potential formulations). The results differ for two reasons: first, $\varepsilon_r\neq\mu_r$; second, the polarization and magnetization are oriented in different directions, namely $x$ and $z$ respectively. When setting $\mu_r=\varepsilon_r$, and when using identical polarization and magnetization directions, the relative differences are identical, as expected. 

\begin{table*}
\centering
\caption{Electrostatic energy simulations, relative errors.\label{tab_ess}}
\renewcommand{\arraystretch}{1.4}
\begin{tabular*}{0.76\linewidth}{@{\extracolsep{\fill}} llllll@{}}

\toprule
Source & Analytical & Trunc. ($\%$) & ABC-1 (ppm) & IABC-3 (ppm) & Shell (ppm)\\ \midrule

\multirow{2}{*}{${P}_{\!p}$} & 
$W_E=115.95239~\mu$J & 
$+0.128$ & 
$+11.1$  &
$-10.9$  &
$-10.1$  
\\

 & 
$W_D=74.807995~\mu$J & 
$-0.198$ & 
$-17.7$  &
$+16.3$  &
$+15.2$  
\\\hline

$\rho_{\!f}$ & 
$W_E=53.718125$~nJ & 
$-9.79$ & 
$+1977$  &
$+17.5$  &
$-23.2$  
\\

$V$ & 
\multirow{2}{*}{$W_E=5.5632503$~nJ} &
$+11.6$ & 
$-2100$  &
$-19.1$  &
$+24.2$  
\\

$\sigma_{\!f}$ & 
 &
$-10.4$ & 
$+2105$  &
$+18.9$  &
$-24.4$  
\\\hline

$\rho_{\!f}\,;\,-\rho_{\!f}$ & 
$W_E=87.468301$~nJ &
$+1.05$ & 
$+8.74$  &
$-6.77$  &
$-14.1$  
\\

$V_0\,;\,-V_0$ & 
$W_E=17.083301$~nJ &
$-1.16$ & 
$-8.78$  &
$+8.11$  &
$+16.5$  
\\
\bottomrule 

\end{tabular*}
\end{table*}

\begin{table*}
\centering
\caption{Magnetostatic energy simulations, relative errors.\label{tab_mss}}
\renewcommand{\arraystretch}{1.4}
\begin{tabular*}{0.85\linewidth}{@{\extracolsep{\fill}} llllll@{}}

\toprule
Source & Analytical & Trunc. ($\%$) & ABC-1 (ppm) & IABC-3 (ppm) & Shell (ppm)\\ \midrule

\multirow{2}{*}{${M}_{\!p}$} & 
$W_H=74.862777$ mJ & 
$+0.187$ & 
$+26.2$  &
$-16.6$  &
$-14.1$  
\\

 & 
$W_B=93.578471$ mJ & 
$-0.150$ & 
$-21.3$  &
$+12.9$  &
$+10.9$  
\\\hline

${J}_{\!f}$ & 
$W_B=6.4168095\cdot10^{-24}$~J & 
$-0.0795$& 
$+20.3$  &
$-1.86$  &
$-3.61$  
\\

${K}_{\!f}$ & 
$W_B=1.3755446\cdot10^{-24}$~J &
$-0.113$ & 
$+29.8$  &
$-1.79$  &
$-4.29$  
\\\hline

${M}_{\!p}\leftarrow\,;\,{M}_{\!p}\leftarrow$ &
$W_H=166.36173$~mJ &
$+0.435$ & 
$-4.12$  &
$-17.0$  &
$-17.9$  
\\

${M}_{\!p}\uparrow\,;\,{M}_{\!p}\uparrow$ & 
$W_H=186.32513$~mJ &
$+0.387$ & 
$-7.44$  &
$-15.4$  &
$-17.3$  
\\

${M}_{\!p}\leftarrow\,;\,{M}_{\!p}\uparrow$ &
$W_H=179.67066$~mJ &
$+0.214$ & 
$+12.7$  &
$-16.8$  &
$-16.1$  
\\
\bottomrule 

\end{tabular*}
\end{table*}

The single-sphere test cases were also simulated in ANSYS Maxwell 2D~\cite{ANSYS} (version 2025 R2) for the IABC method, with the exception of ${P}_{\!p}$, which is not supported by the software. Results were found to be comparable to those reported here, with one discrepancy in the $\vect{M}_p$ test case. 
Energy and coenergy can be obtained using builtin macros ({\small \texttt{<Energy>}} and {\small \texttt{<CoEnergy>}}), where the corresponding post-processing integrals in the fields calculator of the software are:

{\small \texttt{RZIntegrate(Volume(AllObjects), <Energy>)}}

{\small \texttt{RZIntegrate(Volume(AllObjects), <CoEnergy>)}}

\noindent With the above computations, both the reported energy and coenergy are equal (and consistent with the $W_H$ energy), which should not be the case. When calculating the energy integrals without the macros (using $\mu H^2/2$ and $B^2/(2\mu)$ instead), or when performing the same simulation in FEMM, energy and coenergy are different and consistent with the $W_H$ and $W_B$ values in Table~\ref{tab_mss}, respectively. 

\begin{figure*}
\centerline{
	\includegraphics[width=0.49\linewidth,trim=0cm 0cm 0cm 10pt, clip]{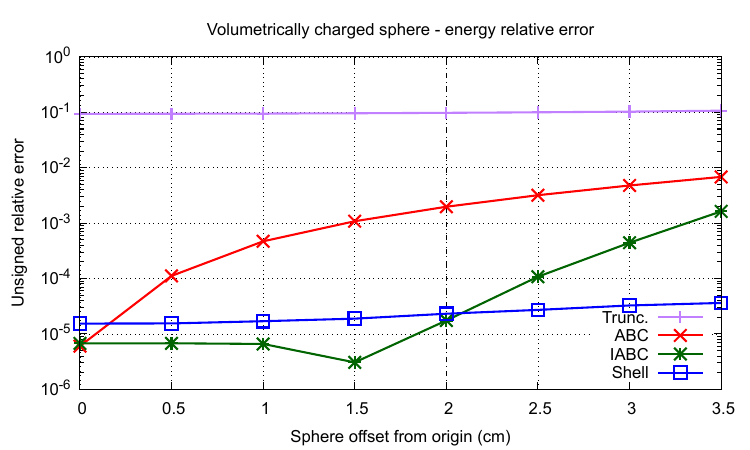}
	\includegraphics[width=0.49\linewidth,trim=0cm 0cm 0cm 10pt, clip]{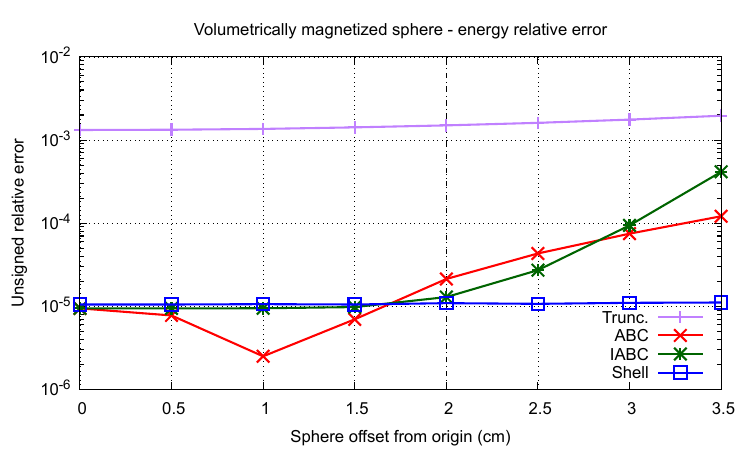}
}
\caption{Effect of the sphere's offset from the origin for the ``$\rho_{\!f}$'' and ``${M}_{\!p}$'' test cases (total energy relative error).\label{figoffset}}
\end{figure*}

The location of the sphere(s) can also have an effect on the performance of the open boundary methods. For best performance, the simulated objects should generally reside as close as possible to the origin, and the boundary should not be placed too close to the objects, especially for the ABC-1 method. In order to evaluate this aspect, two single-sphere test cases are studied with varying displacements of the sphere's center position with respect to the origin. 
Figure~\ref{figoffset} shows that as the sphere is displaced from the origin in the ``$\rho_{\!f}$'' and ``${M}_{\!p}$'' test cases, the shell method behaves the most consistently and generally provides the best accuracy. As is visible in Tables~\ref{tab_ess} and \ref{tab_mss}, the monopolar sources ($\rho_{\!f}$, $V_0$, $\sigma_{\!f}$) are much more challenging for the ABC and truncation methods. This is also visible in Fig.~\ref{figoffset}, when comparing the monopolar case in the left-side graph with the dipolar case in the right-side graph. In these graphs and those in Appendix~\ref{apx_err_conv}, the emphasized vertical grid lines correspond to the results appearing in the tables.

The pronounced increase in relative error observed in the rightmost portions of the IABC-3 curves in Fig.~\ref{figoffset} is mostly a consequence of the boundary regions chosen in the model, and  is caused by the position of the boundary discontinuity at $r=r_b$, which is the closest and most abrupt for the IABC method. For the truncation and ABC methods, the boundary discontinuity is at $r=r_e$ (twice as far), and in the shell method, the spatial distortion at $r=r_b$ is gradual, both of which are more forgiving in this particular experiment.

\subsection{Force and torque}
\label{ssec_fts}

\begin{figure*}
\centerline{\includegraphics[width=0.8\linewidth]{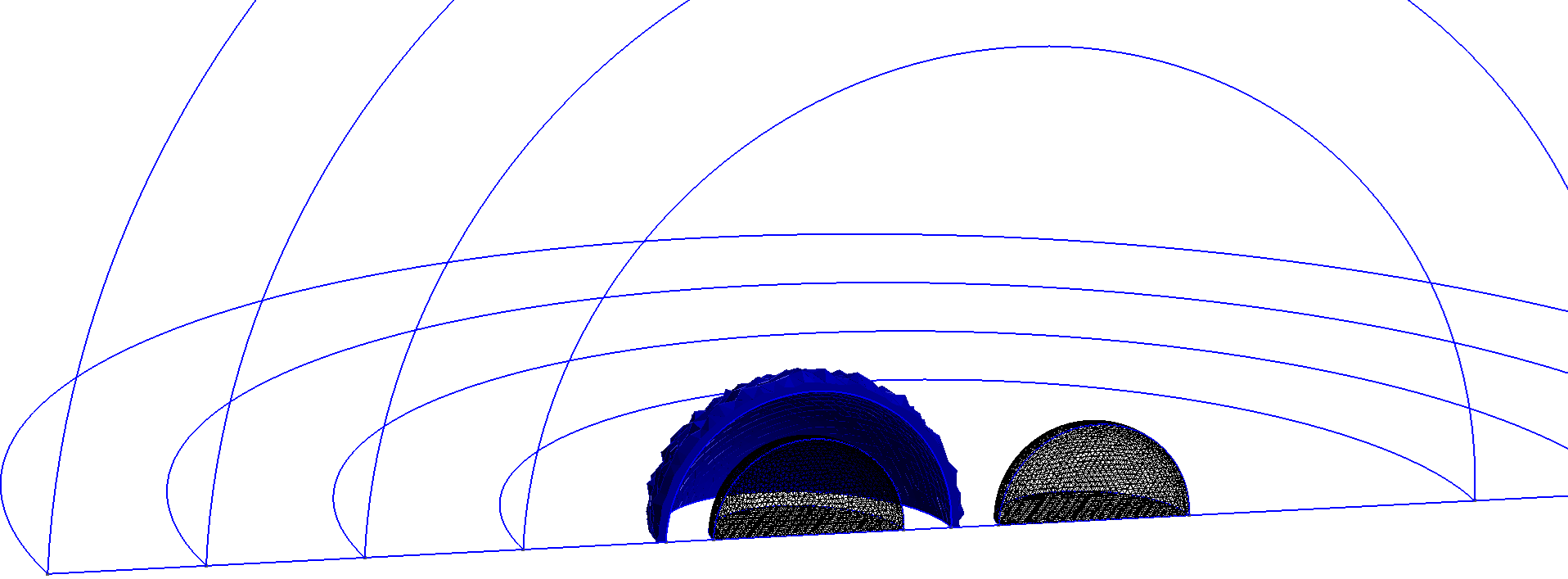}}
\caption{Probe layer elements (dark blue) for computing force and torque on sphere~1 with the Maxwell stress tensor.\label{figprobe}}
\end{figure*}

For added precision in the finite element method, the surface integrals involving the Maxwell stress tensor (Table~\ref{tab_2s}) are implemented as volume integrals on a thin shell surrounding the object on which the force and torque are desired. This one-element thick shell is called a probe layer in the present work. The approach can be described as the (decoupled) eggshell method~\cite{HENROTTE2010} or the weighted stress tensor method~\cite{FEMM_WST}. 
The probe layer for sphere~1 is shown in Fig.~\ref{figprobe}.

Simulation results for the test cases from Table~\ref{tab_2s} are presented in Table~\ref{tab_fts}. Relative differences are greater than in the energy simulations, with the ABC-1 method showing a greater degradation in accuracy compared to the IABC-3 and shell methods. Placing the probe layer directly on sphere~1 results in lower accuracy.

\begin{table*}
\centering
\caption{Force and torque simulations, relative errors.\label{tab_fts}}
\renewcommand{\arraystretch}{1.4}
\begin{tabular*}{0.82\linewidth}{@{\extracolsep{\fill}} llllll@{}}

\toprule
Source & Analytical & Trunc. ($\%$) & ABC-1 (ppm) & IABC-3 (ppm) & Shell (ppm)\\ \midrule

$\rho_{\!f}\,;\,-\rho_{\!f}$ & 
$F=1.1213885~\mu$N &
$+5.58$ & 
$+313$  &
$-17.5$  &
$-40.3$  
\\

$V_0\,;\,-V_0$ & 
$F=353.61901$~nN &
$+2.56$ & 
$+164$  &
$-43.2$  &
$-52.7$  
\\\hline

${M}_{\!p}\leftarrow\,;\,{M}_{\!p}\leftarrow$ &
$F=1.3308938$~N &
$+0.120$ & 
$+282$  &
$-76.2$  &
$+37.8$  
\\

${M}_{\!p}\uparrow\,;\,{M}_{\!p}\uparrow$ & 
\multirow{2}{*}{$F=665.44691$~mN} &
$-0.170$ & 
$-370$  &
$+115$  &
$-22.6$  
\\

${M}_{\!p}\leftarrow\,;\,{M}_{\!p}\uparrow$ &
 &
$+0.107$ & 
$+119$  &
$-66.1$  &
$-60.4$  
\\\hline

${M}_{\!p}\leftarrow\,;\,{M}_{\!p}\uparrow$ &
$\tau=3.3272345$~mN$\cdot$m &
$-10.0$ & 
$+1202$  &
$-123$  &
$-14.6$  
\\
\bottomrule 

\end{tabular*}
\end{table*}

\subsection{Electromagnetic momentum}

Numerical results for the electromagnetic momentum test cases are presented in Table~\ref{tab_ems}, where the IABC and the shell open boundary methods are used. To properly implement the open-space momentum computations with the shell method, the inverse transformation of~(\ref{eqshell}) is used in conjunction with the offset of the sphere's center (this center is also the reference point for angular momentum here). In the case of the IABC method, only the finite space integrals apply since the pointwise correspondence between $\vect{E}$ and $\vect{B}$ does not generally hold in the layers.

The only supplemental constraint applied in the momentum simulations, compared to those outlined in Section~\ref{sec_em}, concerns the ``$\rho_{\!f}\,,\,\vect{M}_p$" test case. The added condition $\varepsilon_r=1$ is set in order to simplify the finite space computation, so that $\rho=\rho_{\!f}$, which avoids both a bound surface charge on the sphere and a weaker internal charge density (because of bound volume charges). 

Since only one quarter of the domain is simulated, the $\vect{M}_p$ and $\boldsymbol{\omega}$ vectors are aligned with one of the axes for symmetry reasons (the $y$-axis is chosen here), as is the $\vect{P}_{\!p}$ vector (the $z$-axis is chosen here). This produces a linear electromagnetic momentum in the $x$-axis for the first test case, and an angular electromagnetic momentum in the $y$-axis for all others. The vector directions can be arbitrary when performing full domain simulations.

\begin{table*}
\centering
\caption{Electromagnetic momentum simulations, relative errors (in ppm).\label{tab_ems}}
\renewcommand{\arraystretch}{1.4}
\begin{tabular*}{0.7\linewidth}{@{\extracolsep{\fill}} lllll@{}}

\toprule
Source & Analytical &IABC-3 (finite) & Shell (finite) & Shell (open)\\ \midrule

${P}_{\!p}\perp{M}_p$ &
$p=1.3761540\cdot10^{-11}$ kg$\frac{\textrm{m}}{\textrm{s}}$ & 
$-8.68$  &
$-7.11$  &
$-7.05$ 
\\[1pt]\hline

$\rho_{\!f}\,,\,{M}_p$ & 
$L=1.4972555\cdot10^{-15}$ kg$\frac{\textrm{m}^2}{\textrm{s}}$ & 
$-3.68$  &
$-5.72$  &
$-12.9$ 
\\

$V_0\,,\,{M}_p$ & 
\multirow{2}{*}{$L=8.2856136\cdot10^{-16}$ kg$\frac{\textrm{m}^2}{\textrm{s}}$} & 
$+5.08$  &
$+122$  &
$+14.7$ 
\\

$\sigma_{\!f}\,,\,{M}_p$ & 
 & 
$-3.47$  &
$-5.51$  &
$-9.92$ 
\\[1pt]\hline

${J}_{\!f}$ & 
$L=1.2833619\cdot10^{-26}$ kg$\frac{\textrm{m}^2}{\textrm{s}}$ & 
$-2.57$  &
$-3.64$  &
$-9.41$ 
\\

${K}_{\!f}$ & 
$L=2.7510892\cdot10^{-27}$ kg$\frac{\textrm{m}^2}{\textrm{s}}$ & 
$-2.77$  &
$-4.30$  &
$-8.73$ 
\\

\bottomrule 

\end{tabular*}
\end{table*}

The momenta results in the first test case of Table~\ref{tab_ems} are calculated with the first of the two finite space integrals given in the first row in Table~\ref{tab_em}. 
All results are globally comparable to the analytical values, with the highest relative error occurring in the finite-space shell result of the ``$V_0\,,\,\vect{M}_p$'' test case. In Table~\ref{tab_ems}, the fact that the values in the two rightmost columns differ from each other in each row indicates that momentum is more numerically demanding to compute compared to energy.

The simulation examples conclude with a visualization of the Poynting vector field $\vect{S}$ associated with a uniformly charged sphere placed near a uniformly magnetized cylindrical magnet. The magnet has a height of 4~cm and a radius of 1~cm, and a uniform coarse mesh is used for visualization purposes. The sphere's volume charge density is $\rho_{\!f}$ and the bar magnet's permanent magnetization is $\vect{M}_p$. The permittivity and permeability of both objects are those of vacuum. 

\begin{figure*}
\begin{tikzpicture}
\node[anchor=north west] at (-3.2,0) {\includegraphics[width=0.98\linewidth]{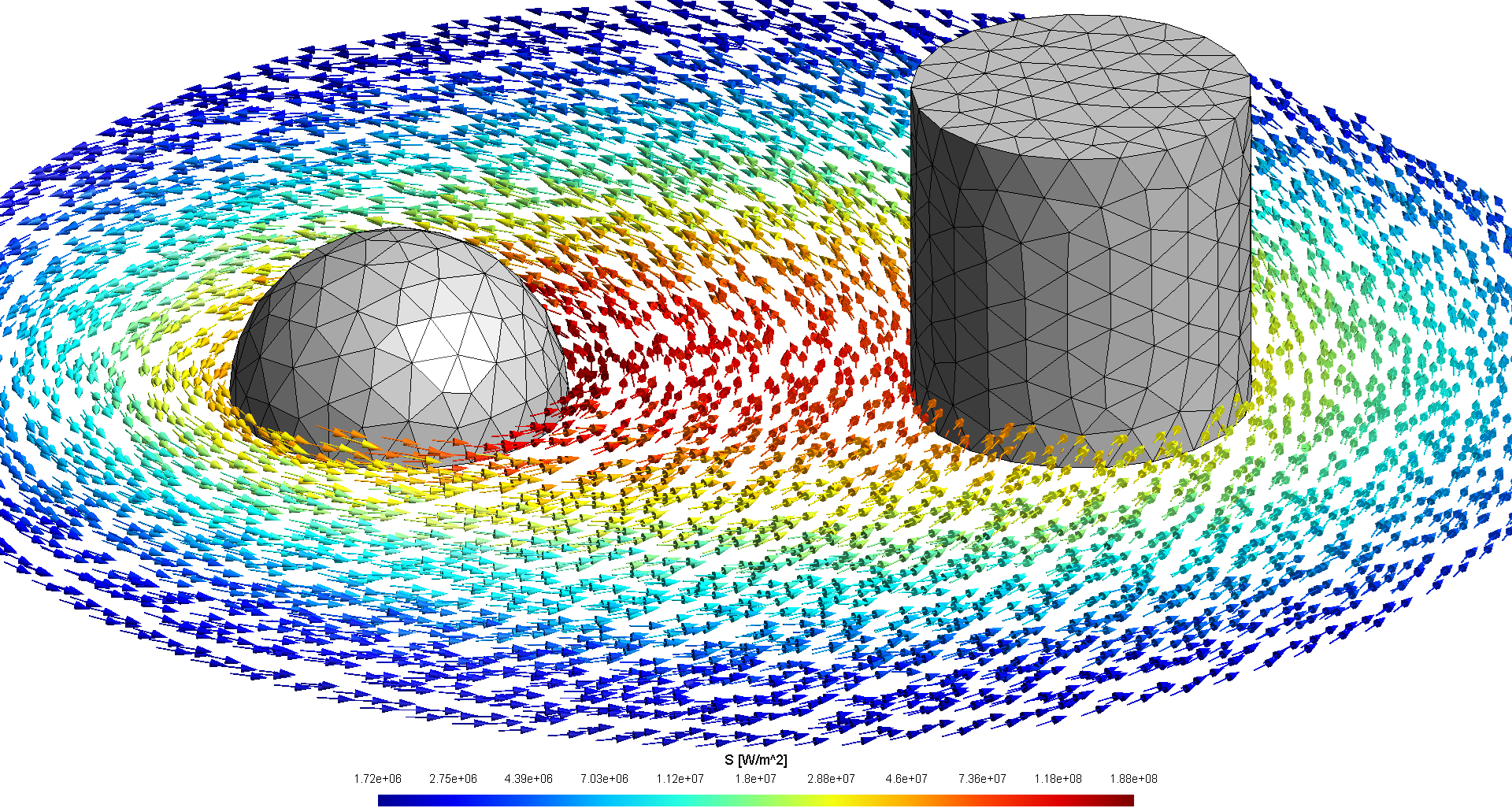}};

\def\scl{1.0}

\coordinate (Q) at (-.1,0);
\coordinate (MBL) at (0.05,-0.35);
\coordinate (MTR) at (0.17,0.35);

\filldraw [draw=white,fill=white] (-3.3,-1.5) rectangle ++(6.4,2.75);

\draw[fill=black] (Q) circle (0.05);
\draw[anchor=south] (-0.1,-0.01) node[] {\scalebox{0.6}{$q$}};

\draw [arrows = {-Stealth[inset=1.5pt,width=6pt, length=9pt]}] (Q) to ++(3,1);
\draw[anchor=south] (-3.0,0.35) node[] {\scalebox{\scl}{$\vect{E}$}};
\draw [arrows = {-Stealth[inset=1.5pt,width=6pt, length=9pt]}] (Q) to ++(3,-1);
\draw [arrows = {-Stealth[inset=1.5pt,width=6pt, length=9pt]}] (Q) to ++(-3.2,-1);
\draw [arrows = {-Stealth[inset=1.5pt,width=6pt, length=9pt]}] (Q) to ++(-3.2,1);

\filldraw [draw=black,fill=white] (MBL) rectangle (MTR);
\draw[anchor=west] (0.11,0.4) node[] {\scalebox{0.6}{N}};
\draw[anchor=west] (0.11,-0.4) node[] {\scalebox{0.6}{S}};

\def\a{1.3} 
\def\b{1.2} 
\draw[] (0.1,0.35) arc[start angle=17.1, end angle=342.9, x radius=\a, y radius=\b];
\draw [arrows = {-Stealth[inset=1.5pt,width=6pt, length=9pt]}] (-2.435,0) to ++(0,-0.2);
\draw[] (0.12,0.35) arc[start angle=17.1, end angle=342.9, x radius=-\a, y radius=\b];
\draw [arrows = {-Stealth[inset=1.5pt,width=6pt, length=9pt]}] (2.655,0) to ++(0,-0.2);
\draw[anchor=south] (-2.75,-0.3) node[] {\scalebox{\scl}{$\vect{B}$}};
\filldraw [draw=white,fill=white] (-0.6,-1.0) rectangle ++(0.4,0.05);
\filldraw [draw=white,fill=white] (0.2,-0.975) rectangle ++(0.4,0.05);

\def\c{1.8} 
\def\d{1} 
\draw[dashed] (0,0) ++(\c,0) arc[start angle=0, end angle=180, x radius=\c, y radius=\d];
\draw[] (0,0) ++(-\c,0) arc[start angle=180, end angle=360, x radius=\c, y radius=\d];
\draw [arrows = {-Stealth[inset=1.5pt,width=6pt, length=9pt]}] (0.0,-\d) to ++(0.2,0);
\draw[anchor=north] (0.1,-\d) node[] {\scalebox{\scl}{$\vect{S}$}};


;
\end{tikzpicture}
\caption{Poynting vector field $\vect{S}=c^2\vect{g}$ in the one-element-thick layer above the plane of symmetry, outside of a charged sphere and a bar magnet (coarse mesh, top hemisphere of interior domain). Top-left diagram is the author's rendition of Fig. II-27-6 from Feynman et al.~\cite{FEYNMAN1964}.\label{figFeynman}}
\end{figure*}
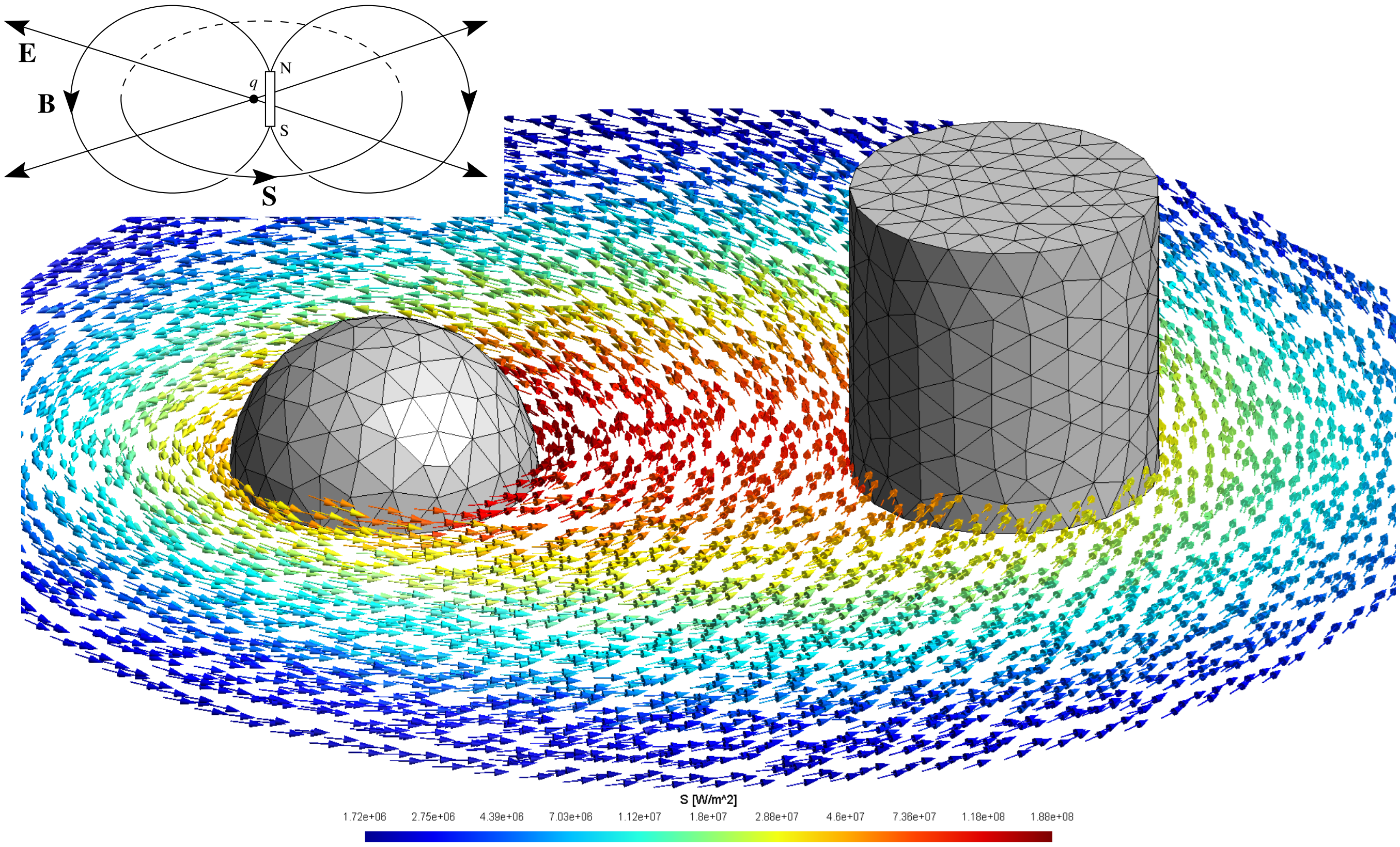

Figure~\ref{figFeynman} reveals a circular flow of energy in the fields outside of the objects. By a simple change of scale, this illustration also represents the momentum density $\vect{g}$, because $\vect{S}=c^2\vect{g}$. 
For a point charge in the presence of a magnet, momentum approximations exist~\cite{FURRY1969}, and the Poynting vector has been previously sketched~\cite{FEYNMAN1964}. 
Even without exact expressions for comparison, especially when a sphere is used rather than a point charge, high confidence in the accuracy of this simulation is nonetheless reasonable given the verifications performed throughout this section. 

\section{Conclusion}

A library of test cases based on the sphere has been presented for verifying the correctness of open boundary simulations in electrostatics and magnetostatics. Analytical expressions for energy, force, torque and momentum were developed for use as benchmarking metrics, with generalizations beyond those normally reported in the literature. These metrics can be particularly beneficial when the computation of $L_2$ error norms is either cumbersome or infeasible in a given software package.

Numerical experiments were performed to illustrate the application of the test cases to four different open boundary methods. 
The truncation and ABC-1 methods performed poorly for uncentered monopolar sources, as could be expected, while the IABC-3 and shell methods performed substantially better. For dipolar sources, the ABC-1 method was broadly comparable to the IABC-3 and shell methods in the energy simulations, although to a lesser extent in the force and torque experiments. 
Once verified, the models were used to visualize the Poynting vector field in a configuration for which no analytical result is available. 

As demonstrated in one experiment with the ANSYS Maxwell software, test cases with known results helped ensure the proper behavior of post-processing calculations. 
Whether in custom simulation codes or commercial software, verification is essential for ensuring reliable results. The set of analytical expressions presented in this work is intended to facilitate the verification of open boundary simulations in electrostatics and magnetostatics for computational physics and engineering applications.



\section*{Acknowledgements}

The author would like to thank Danielle Bouthot, Alain H\'enault, J\'er\^ome Quintin, and the anonymous reviewers for suggestions that helped improve the paper.

\section*{Code and data availability}

Simulation files are made available in a public GitHub repository~\cite{SIMFILES} and all numerical experiments are reproducible with the ONELAB open source software~\cite{ONELAB}.

\appendix

\section{Error evaluation and mesh convergence}
\label{apx_err_conv}

The normalized $L_2$ error norms~\cite{FISH2007} of the electric and magnetic fields are presented in Tables~\ref{tab_ess2} and~\ref{tab_mss2} for the test cases appearing in Tables~\ref{tab_ess} and~\ref{tab_mss} respectively. 
For the configuration involving two conducting spheres, the error is not evaluated because the corresponding electric field expression requires taking the gradient (in bi-spherical coordinates) of a potential expressed as an infinite sum over Legendre polynomials~\cite{MORSE1953}, which does not admit a simple analytical expression.

The normalized $L_2$ error norm in the electric field is given below, where $\vect{E}_\text{ex}$  is the field obtained using exact expressions and  $\vect{E}_\text{sim}$ is obtained by simulation. The calculation is performed in the $r<r_b$ region (called $\Omega_b$), where the field is expected to be directly comparable between all boundary methods.
\begin{equation}
\label{eql2}
\bar{\text{e}}_{L_2}=\frac{\|\vect{E}(\vect{r})_\text{ex} - \vect{E}(\vect{r})_\text{sim}\|_{L_2}}{\|\vect{E}(\vect{r})_\text{ex}\|_{L_2}}=
\frac{\left(\int_{\Omega_b}\|\vect{E}(\vect{r})_\text{ex} - \vect{E}(\vect{r})_\text{sim}\|^2\,d\Omega_b\right)^{1/2}}{\left(\int_{\Omega_b}\|\vect{E}(\vect{r})_\text{ex}\|^2\,d\Omega_b\right)^{1/2}}
\end{equation}
A similar calculation can be performed with the magnetic field. A higher number of integration points is used for these post processing integrals to ensure sufficient accuracy.

The normalized $L_2$ error norms in Tables~\ref{tab_ess2} and~\ref{tab_mss2} are noticeably greater than the relative errors in energy in Tables~\ref{tab_ess} and~\ref{tab_mss}. Relative errors in energy involve the difference of squared norms of fields, whereas the normalized $L_2$ error norms involve the squared norm of the difference of fields, after which the square root is taken. For a randomly distributed field error, the square root in (\ref{eql2}) is the main reason why both metrics differ in magnitude and should be taken into consideration when assessing simulation accuracy.

\begin{table}
\centering
\caption{Electrostatic simulations, normalized $L_2$ error norms of~$\vect{E}$ (in $\%$).\label{tab_ess2}}
\renewcommand{\arraystretch}{1.4}
\begin{tabular*}{0.9\linewidth}{@{\extracolsep{\fill}} lllll@{}}

\toprule
Source & Trunc. & ABC-1 & IABC-3 & Shell\\ \midrule

${P}_{\!p}$ & 
$1.3613$ & 
$0.3839$  &
$0.3827$  &
$0.3826$  
\\

$\rho_{\!f}$ & 
$1.4750$ & 
$0.7700$  &
$0.2541$  &
$0.2456$  
\\

$V$ & 
$11.5430$ & 
$0.7860$  &
$0.2576$  &
$0.2490$  
\\

$\sigma_{\!f}$ & 
$1.4923$ & 
$0.7791$  &
$0.2575$  &
$0.2489$  
\\

$\rho_{\!f}\,;\,-\rho_{\!f}$ & 
$2.9964$ & 
$0.2272$  &
$0.2263$  &
$0.2262$  
\\

$V_0\,;\,-V_0$ & 
N.A. & 
N.A. &
N.A. &
N.A.
\\
\bottomrule 

\end{tabular*}
\end{table}

\begin{table}
\centering
\caption{Magnetostatic simulations, normalized $L_2$ error norms of~$\vect{B}$ (in $\%$).\label{tab_mss2}}
\renewcommand{\arraystretch}{1.4}
\begin{tabular*}{0.9\linewidth}{@{\extracolsep{\fill}} lllll@{}}

\toprule
Source & Trunc. & ABC-1 & IABC-3 & Shell\\ \midrule

${M}_{\!p}$ & 
$0.9683$ & 
$0.2677$  &
$0.2666$  &
$0.2664$  
\\

${J}_{\!f}$ & 
$0.7974$& 
$0.1882$  &
$0.1856$  &
$0.1854$  
\\

${K}_{\!f}$ & 
$0.9482$ & 
$0.2001$  &
$0.1966$  &
$0.1964$  
\\

${M}_{\!p}\leftarrow\,;\,{M}_{\!p}\leftarrow$ &
$1.2970$ & 
$0.2508$  &
$0.2507$  &
$0.2506$  
\\

${M}_{\!p}\uparrow\,;\,{M}_{\!p}\uparrow$ & 
$1.3348$ & 
$0.2672$  &
$0.2671$  &
$0.2670$  
\\

${M}_{\!p}\leftarrow\,;\,{M}_{\!p}\uparrow$ &
$0.9586$ & 
$0.2630$  &
$0.2623$  &
$0.2622$  
\\
\bottomrule 

\end{tabular*}
\end{table}

\begin{figure*}
\centerline{
	\includegraphics[width=0.49\linewidth,trim=0cm 0cm 0cm 10pt, clip]{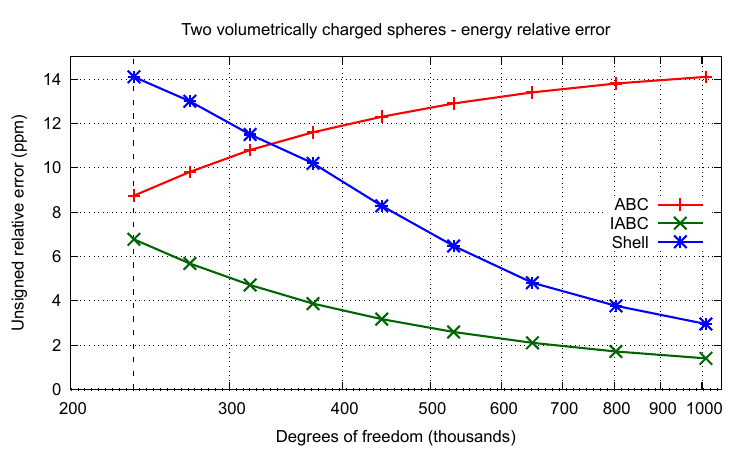}
	\includegraphics[width=0.49\linewidth,trim=0cm 0cm 0cm 10pt, clip]{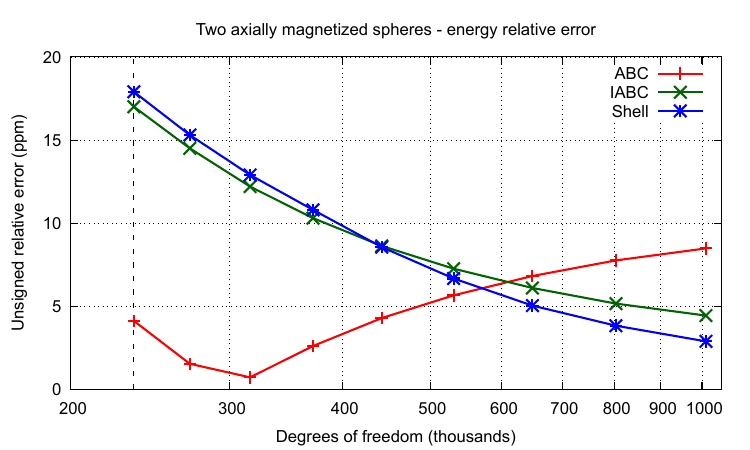}
}
\caption{Mesh convergence analysis for the ``$\rho_{\!f}\,;\,-\rho_{\!f}$'' and ``${M}_{\!p}\leftarrow\,;\,{M}_{\!p}\leftarrow$'' test cases (total energy relative error).\label{figconv}}
\end{figure*}

Mesh convergence is evaluated for the two test cases having the lowest relative errors in Tables~\ref{tab_ess} and~\ref{tab_mss}, as they could be considered the most suspicious. The smallest relative error in the electrostatic energy results corresponds to the IABC-3 entry in the ``$\rho_{\!f}\,;\,-\rho_{\!f}$'' test case. The smallest relative error in the magnetostatic energy results corresponds to the ABC-1 entry in the ``${M}_{\!p}\leftarrow\,;\,{M}_{\!p}\leftarrow$'' test case. 
Figure~\ref{figconv} shows that in both instances, the relative error actually increases for the ABC-1 boundary type as the mesh is refined, while the opposite is true for the IABC-3 and shell boundary types. Since the shell is an exact method, it can be expected that the relative error would tend toward zero as the mesh is refined, which is not the case for the two other methods (although with the IABC-3 method, the relative error would tend toward zero for an origin-centered sphere).
The ABC-1 energy values used in the graphs are obtained with the finite space integrals, while for the IABC-3 and shell methods, both the finite and open space integrals yield identical energies to eight significant digits.

\begin{figure*}
\centerline{
	\includegraphics[width=0.49\linewidth,trim=0cm 0cm 0cm 10pt, clip]{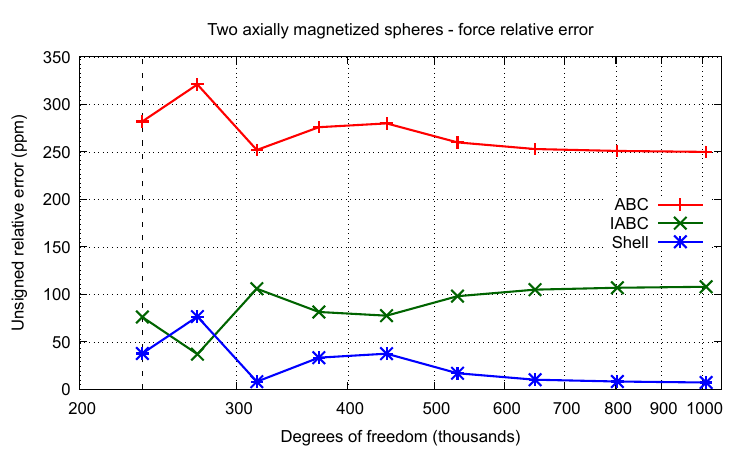}
	\includegraphics[width=0.49\linewidth,trim=0cm 0cm 0cm 10pt, clip]{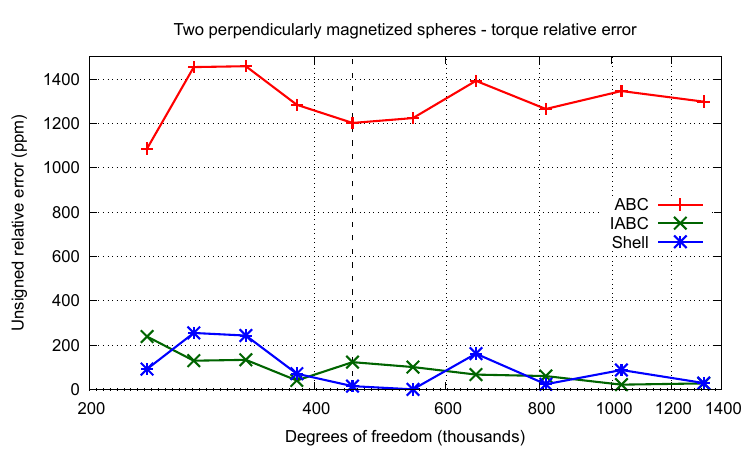}
}
\caption{Mesh convergence analysis for the ``${M}_{\!p}\leftarrow\,;\,{M}_{\!p}\leftarrow$'' and ``${M}_{\!p}\leftarrow\,;\,{M}_{\!p}\uparrow$'' test cases (force and torque relative error).\label{figconv2}}
\end{figure*}

\begin{figure*}
\centerline{
	\includegraphics[width=0.49\linewidth,trim=0cm 0cm 0cm 10pt, clip]{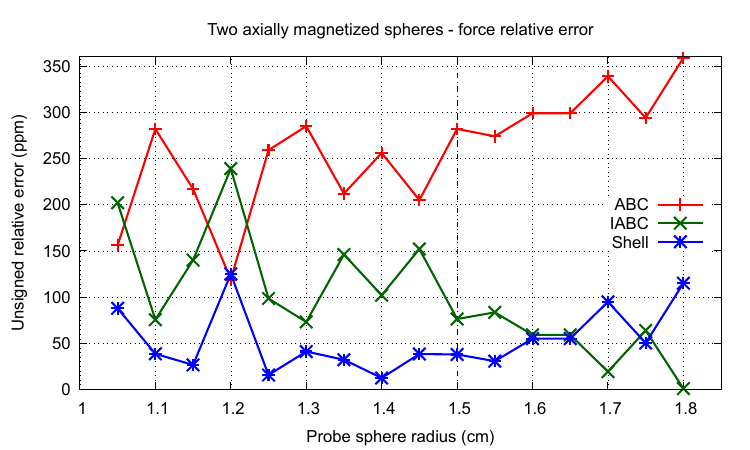}
	\includegraphics[width=0.49\linewidth,trim=0cm 0cm 0cm 10pt, clip]{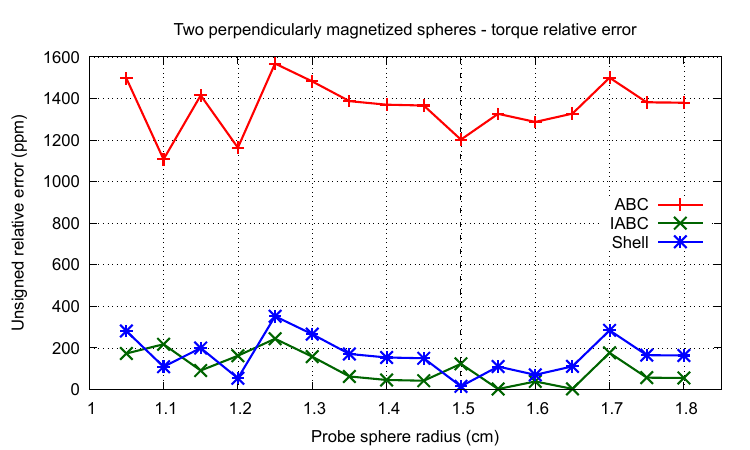}
}
\caption{Probe radius stability analysis for the ``${M}_{\!p}\leftarrow\,;\,{M}_{\!p}\leftarrow$'' and ``${M}_{\!p}\leftarrow\,;\,{M}_{\!p}\uparrow$'' test cases (force and torque relative error).\label{figconv3}}
\end{figure*}

Mesh convergence is evaluated in Fig.~\ref{figconv2} for the force and torque of the ``${M}_{\!p}\leftarrow\,;\,{M}_{\!p}\leftarrow$'' and ``${M}_{\!p}\leftarrow\,;\,{M}_{\!p}\uparrow$'' test cases, revealing that postprocessing with the probe layer can be less stable, although convergence is still generally good. 
A final analysis is presented in Fig.~\ref{figconv3} for the same two test cases, this time varying the radius of the probe sphere. The greatest stability occurs roughly near $1.5$~cm, where the probe layer passes through the midpoint of the two magnetized spheres (as shown in Fig.~\ref{figprobe}). The relative errors are noticeably higher for force and torque than for energy, which is also a consequence of computing integrals in the probe layer. 

\bibliographystyle{unsrtnat}
\bibliography{article}

\end{document}